\documentclass[twoside]{article}
\usepackage{amsmath,amssymb}
\DeclareMathOperator{\diag}{diag}

\usepackage{epsfig}
\usepackage{epstopdf} % For pdfLatex to use .eps
\usepackage{float}   % Improves float objects handling

\usepackage[braket,qm]{qcircuit}
\newcommand{\rgate}[1]{*=<2.1em>[Fo]{#1} \qw}
\newcommand\floor[1]{\lfloor#1\rfloor}
\newcommand\ceil[1]{\lceil#1\rceil}

\usepackage[boxed]{algorithm2e}

\usepackage{enumerate}
\usepackage{cite}
\usepackage[title]{appendix}

\usepackage{amsfonts}
\usepackage{multirow}
\numberwithin{equation}{section} %section #s

\begin{document}
\setlength{\textheight}{8.0truein}    %FOR 2ND PAGE ONWARDS

%\runninghead{A Novel Finite Fractional Fourier Transform and its Arithmetic Quantum Circuits}
%{E. Floratos and A. Pavlidis}

%\normalsize\textlineskip
\thispagestyle{empty}
\setcounter{page}{1}

%\copyrightheading{Vol.}{No.}{Year}{Page Nos.}
%\copyrightheading{XX}{Y\&Y}{2018}{ZZZZ--ZZZZ}

%\vspace*{0.88truein}

\centerline{\bf
%%%%%%%%%%%%%%%%%%%%%
%Put in titiles here
%%%%%%%%%%%%%%%%%%%%%
A NOVEL FINITE FRACTIONAL   FOURIER TRANSFORM}
\vspace*{0.035truein}
\centerline{\bf  AND ITS QUANTUM CIRCUIT IMPLEMENTATION ON QUDITS}
\vspace*{0.37truein}
%%%%%%%%%%%%%%%%%%%%%%%%%%%%%%%%%%%%
%put authors' name and address here
%%%%%%%%%%%%%%%%%%%%%%%%%%%%%%%%%%%%
\centerline{\footnotesize 
EMMANUEL FLORATOS}
\vspace*{0.015truein}
\centerline{\footnotesize\it Research office of Mathematical Physics and Quantum Information, Academy of Athens}
\baselineskip=10pt
\centerline{\footnotesize\it Panepistimiou 28, GR 106 79, Athens,Greece}
\baselineskip=10pt
\centerline{\footnotesize\it Department of Physics, National and Kapodistrian University of Athens}
\baselineskip=10pt
\centerline{\footnotesize\it Panepistimiopolis, Ilissia, GR 157 84, Athens, Greece}
\baselineskip=10pt
\centerline{\footnotesize\it Institute of Nuclear and Particle  Physics, N.C.S.R. Demokritos}
\baselineskip=10pt
\centerline{\footnotesize\it 27, Neapoleos str., Agia Paraskevi, GR 153 41, Athens, Greece}
\baselineskip=10pt
\centerline{\footnotesize\it e-mail: mflorato@phys.uoa.gr}
\vspace*{10pt}

\centerline{\footnotesize
ARCHIMEDES PAVLIDIS}
\vspace*{0.015truein}
\centerline{\footnotesize\it Department of Informatics, University of Piraeus}
\baselineskip=10pt
\centerline{\footnotesize\it 80, Karaoli \& Dimitriou str., GR 185 34, Piraeus, Greece}

\centerline{\footnotesize\it e-mail: adp@unipi.gr}

\vspace*{0.225truein}
%% %%\publisher{Month1 Day1, 201X}{Month2 Day2, 201X}

\vspace*{0.21truein}

%% \abstracts{first paragraph}{second paragraph}{third paragraph}
%% If there is only one paragraph, just keep the second and third empty 
%% like the following one 
%\abstracts{
%%%%%%%%%%%%%%%%%%%%
% put abstract here
%%%%%%%%%%%%%%%%%%%%

\vspace*{10pt}

%\keywords{fractional Fourier transform, QFT, quantum multiplier, quantum diagonal operator, qudit}
\vspace*{3pt}

\vspace*{10pt}

\abstract{

We present a new number theoretic definition of  discrete fractional Fourier transform (DFrFT) .
In this approach the DFrFT is defined as the $N \times N$ dimensional unitary representation of the generator  of the arithmetic  rotational   group  $SO_{2}[\mathbb{Z}_N]$, which  is the finite set of  $\bmod N$ integer, $2\times 2$ matrices acting on the points of the discrete toroidal phase space lattice $\mathbb{Z}_N \times \mathbb{Z}_N$,  preserving the euclidean distance $\bmod N$.

Using known factorization properties of $SO_{2}[\mathbb{Z}_N]$ for any integer $N$ 
into products of $SO_{2}[\mathbb{Z}_{p^{n}}]$'s where $p^n$ are the prime power factors of $N$
and $n=1,2,\ldots$, it is enough to study the arithmetic rotation group  $SO_{2}[\mathbb{Z}_{p^n}]$ which is abelian and cyclic. We show that we  can find always  an appropriate power of the generator of $SO_{2}[\mathbb{Z}_{p^n}]$, which produces  the  rotation by $90$ degrees whose the unitary representation is the discrete  Fourier transform (DFT).

We construct explicitly, using techniques of the Finite Quantum Mechanics (FQM), the $p^n$ dimensional unitary matrix representation of the group $SO_{2}[\mathbb{Z}_{p^n}]$ and especially we work out in detail the one which corresponds to the generator. This is our definition of the arithmetic fractional Fourier transform (AFrFT).

Following this definition, we proceed  to   the construction of efficient quantum circuits for  the AFrFT, on sets of $n$ $p$-dimensional qudits with $p$ a prime integer, by introducing novel quantum subcircuits for diagonal  operators with quadratic phases as well as new quantum subcircuits for multipliers by a constant. The quantum subcircuits that we introduce provide a set capable  to construct quantum circuits for any element of a more general group, the group of Linear Canonical Transformations (LCT),  $SL_{2}[\mathbb{Z}_N]$ of the toroidal phase space lattice. As a byproduct, extensions of the diagonal and multiplier quantum circuits for both the qudit and qubit case are given, which are useful alone in various applications. Also, we analyze the depth, width and gate complexity of the efficient  AFrFT quantum circuit and we estimate its gate complexity which is of the order $O(n^2)$, its  depth which is of the order $O(n)$ with depth $n$, while at the same time it has a structure permitting local interactions between the qudits. 
}

\section{Introduction}\label{sec:Introduction} 
One of the most important and historically influential algorithm in the field of quantum computation is the Shor's Factorization algorithm of very large integers \cite{Shor:1994}. The exponential improvement of computational  complexity of this algorithm, compared to the existing classical ones, is mainly due to the Quantum Fourier Transform (QFT) and its quantum circuit.

In this work we extend the QFT and its quantum circuit to the Quantum arithmetic fractional Fourier transform (QAFrFT) for multiqudit quantum systems. Before doing that we shall  review some basic facts which will help us to introduce the new definition of the discrete fractional Fourier transform which we call arithmetic fractional Fourier transform (AFrFT). This definition differs from the standard one which does not have group theoretic origin \cite{Ozaktas:2001,Saxena:2005}.

The standard fractional Fourier tranform (FrFT) has been introduced as a fractional power of the  Fourier transform  and although it  has been studied and  applied in  various scientific fields (harmonic analysis and differential equations, number theory, quantum mechanics, tomography, signal processing) \cite{Ozaktas:2001,Saxena:2005,Condon:1937,Bargmann:1961,Moshinsky:1971,Bruijn:1973,Namias:1980,Mustard:1991,Lohmann:1993,Almeida:1994,Huang:2011,Pegard:2011,Somma:2016,Grafe:2017,Tschernig:2018,Sejdić:2011,Healy:2016}, its main area of application is that of optics, optoelectronics and telecommunications \cite{Ozaktas:2001}.

Our interest focuses in the proposal of a Quantum FrFT based on  quantum mechanics, where  a mathematical  construction with a  precise geometrical interpretation exists using the representation theory of the LCT \cite{Condon:1937,Bargmann:1961,Namias:1980,Healy:2016}. To be concrete we remind the reader that the initial construction of  the FrFT has taken place in the framework of Quantum Mechanics (QM) and it  is cosidered as a fractional power of the $90^\circ$ rotation, which is exactly the Fourier transform (FT), in the $2-$dimensional   phase space plane of the position  and momentum of a quantum  harmonic oscillator \cite{Condon:1937,Bargmann:1961,Moshinsky:1971,Bruijn:1973,Namias:1980}.

We would like to recall at this point  that in QM the FT has a deep meaning of relating the  dual aspects of particles and waves at the atomic scale. Its classical analogue is essentially a prisme analysing waves into their harmonic components. In the classical theory there is no relation between particles and waves. The striking and revolutionay  consequence of this quantum mechanical, particle-wave duality relation, is that in QM we cannot measure simultaneously position and momentum and that means that the correspoding dual variables are not numbers but operators (matrices) with well defined commutation relations. These are called the Heisenberg canonical commutation relations, which lead to the reknown uncertainty relations \cite{Schwinger:2013}.

The good news is  that  the continuous FrFT can be discretized and reduced to finite size matrices, because of the existence of an exact framework of the finite and discrete Quantum Mechanics and of  the representation theory of the discrete and finite LCT \cite{Hannay:1980,Balian:1986,Floratos:1989,Athanasiu:1994,Athanasiu:1996,Athanasiu:1998,Vourdas:2004}.

Passing over to the discrete and finite domain of these dual variables proves to be  a tricky business and for this reason we have to  introduce the necessary  basic mathematical formalism of the FQM, in Section \ref{sec:FQMandFrFT} \cite{Balian:1986,Floratos:1989,Athanasiu:1994,Athanasiu:1996}. In this framework the problem of defining the FrFT is reduced to study  the group of LCT, $SL_{2}[\mathbb{Z}_N]$,  which leaves invariant the area of the phase space lattice $\mathbb{Z}_N \times \mathbb{Z}_N$.
This group is defined as the set of $2\times2$ integer  matrices with multiplication operation $\bmod N$ and determinant equal to $1 \bmod N$ and has been studied thoroughly in mathematics and physics \cite{Hannay:1980,Balian:1986,Floratos:1989,Athanasiu:1994,Athanasiu:1996,Athanasiu:1998}. The DFT corresponds to an element of the  abelian subgroup of discrete rotations, which preserves the Euclidean distance  $\bmod N$ and which is denoted by $SO_{2}[\mathbb{Z}_N]$ \cite{Athanasiu:1994,Athanasiu:1996}. 
For $N=p^n$, where $p$ is an odd prime and $n=,1,2,\ldots$ the order of  $SO_{2}[\mathbb{Z}_{p^n}]$, which in this case is cyclic, is known to be $p^{(n-1)}(p+1)$ for $p=3\bmod 4$ and $p^{(n-1)}(p-1)$ for $p=1\bmod 4$.
The problem to find a generator of this group is hard and it must be solved   by random search\cite{Athanasiu:1994,Athanasiu:1996}.

The order of the group  $SO_{2}[\mathbb{Z}_N]$, can be found for arbitray integer $N$ using its prime factorization and the Chinese remainder theorem \cite{Athanasiu:1998}. One notes  that for any $p$  prime and $n=1,2,\ldots$ the order of the group is divisible by $4$, so there is a power $k$ of the generator with order $4$, which  corresponds to the DFT with radix $p^n$. All the powers of the generator  define the possible  fractional powers of the DFT.

We take as our definition of the AFrFT  the unitary matrix corresponding to the generator  of the rotation group $\bmod N$, $SO_{2}[\mathbb{Z}_N]$ according to the  rules of FQM. Armed with this definition and the construction of a specific (Weil) unitary representation of the finite LCT and thus of the subgoup $SO_{2}[\mathbb{Z}_p^n]$, which we present in Section \ref{sec:FQMandFrFT}, we find the $p^n \times p^n$ unitary matrix representation of the AFrFT \cite{Athanasiu:1996}. All the elements of this unitary matrix are $n-$th roots of unity.

The plan of the paper goes as follows:

In section  \ref{sec:LCT} we introduce the various subgroups of $SL_{2}[\mathbb{Z}_N]$ in terms of which any of its abelian element can be factorized and we define the rotational subgroup $SO_{2}[\mathbb{Z}_{p^{n}}]$ as well as we study its properties. In particular we decompose the generator of the rotation group $\bmod N$ into left translations, diagonal and right translation group elements.

In section \ref{sec:FQMandFrFT} we construct explicitly the unitary metaplectic representation $SO_{2}[\mathbb{Z}_{p^{n}}]$ of dimensionality $p^n$ and the corresponding matrix for the generator of this abelian group, provides our definition of the  AFrFT. The need for new diagonal multiqudit subcircuit with quadratic phases becomes obvious after using the above decomposition. We discuss the different geometrical meaning of AFrFT, which is consistent with the rotation group $\bmod p^n$  of the toroidal discrete phase space, in contrast with the standard DFrFT which violates explicitly the rotation invariance in the discrete phase space \cite{Ozaktas:2001}.

In Section    \ref{sec:QFTQudits} we introduce the necessary basic qudit gates used in our design, as well as the QFT circuit on qudits with local interactions.

In Section  \ref{sec:MultQudits} we propose an in-place (without any ancilla) modulo $p^n$ multiplier of linear depth, quadratic quantum cost and local interactions between the qudits. 

In Section \ref{sec:DiagQudits} we propose circuits which perform as diagonal operators of qudratic phases. The characteristic of these circuits is that they are ancilla-free, they have linear depth, quadratic quantum cost and local interactions like the modulo multiplier.

In Section  \ref{sec:QFrFTcircuit} we combine the previous circuits to build the quantum multiqudit circuit for the AFrFT (QAFrFT) and give the estimation of the depth and the quantum cost of the whole circuit.

in Section \ref{sec:Variations} we extend the modulo multiplier to a normal constant multiplier for any constant, which can operate on multilevel qudits, as well as its adaptation to the two-dimensional qubits case. Also, we show that the diagonal operator quantum circuit can be directly adapted to the qubit case.

We conclude with  our results, open problems and possible applications in Section \ref{sec:Conclusions}.

\section{The LCT group $SL_{2}[\mathbb{Z}_N]$}\label{sec:LCT}

In this Section we discuss the discrete LCT group $SL_{2}[\mathbb{Z}_N]$ and we study certain of its subgroups, especially the discrete and finite rotation group $SO_{2}(\mathbb{Z}_N)$. This special subgroup will help us to introduce the new version of the discrete and finite AFrFT. We begin with the basic definitions. The set $\mathbb{Z}_N=\{0,1,2,\ldots, N-1\}$ is the set of residue classes for integers $\bmod N$, $N$ an integer, and it is equipped with the algebraic operations of addition and multiplication $\bmod N$.

The two dimensional lattice torus $\mathbb{T}_{N}^{2}=\mathbb{Z}_N \times \mathbb{Z}_N$ is the discrete analogue of the phase space of classical mechanics torus $\mathbb{T}^{2}=\mathbb{R} \times \mathbb{R}$ with coordinates $(q,p)\in \mathbb{T}^{2}$, the position and momentum of one point particle.

The group of continuous LCT, is the group of $2 \times 2$ real matrices $SL_{2}(\mathbb{R})$ with determinant equal to one which preserve the area of the phase space. Its discrete analogue is the group $SL_{2}[\mathbb{Z}_N]$ acting on the discrete torus $(q,p)\in \mathbb{T}_{N}^{2}$, as follows:

\begin{equation} 
(q,p)_{m+1} = (q,p)_{m} A \bmod N,\quad m=0,1,...
\label{SLaction}
\end{equation}

\begin{equation} 
A=\left(\begin{array}{cc}a&b\\c&d\end{array}\right) \in SL_{2}[\mathbb{Z}_N], \quad a,b,c,d \in \mathbb{Z}_N
\label{Adef}
\end{equation}
and $det(A)=1 \bmod N$.

The index $m=0,1,\ldots$ represents the iteration or discrete time of the motion of a point particle initially  ($m=0$) at $(q,p)_{0}=(q_{0},p_{0})$. The $\bmod N$ operation guarantees that the particle during its motion remains always inside the torus $\mathbb{T}_N^{2}$. Since the number of the points in $\mathbb{T}_N^{2}$ is finite, $N^2$, any element $A \in SL_{2}[\mathbb{Z}_N]$ has a finite period, i.e. there is a smallest integer $T_{A}(N)$ such that 

\begin{equation}
A^{T_{A}(N)}=I_{2\times 2} \bmod N.
\label{Aperiod}
\end{equation}
The period $T_{A}(N)$ of $A$ is an erratic function of $N$. As $N$ grows we find smaller or larger periods in a random way \cite{Dyson:1992}.  

There are various interesting subgroups of $SL_{2}[\mathbb{Z}_N]$. The left and right abelian translation groups $L$ and $R$ with elements $L(a)$, $R(a)$, respectively:

\begin{equation}
L(a)=\left(\begin{array}{cc}1&0\\a&1\end{array}\right),  \quad a \in \mathbb{Z}_N
\label{Lsubgroup}
\end{equation}

\begin{equation}
R(a)=\left(\begin{array}{cc}1&a\\0&1\end{array}\right), \quad a \in \mathbb{Z}_N
\label{Rsubgroup}
\end{equation}
with generators 

\begin{equation}
g_{L}=\left(\begin{array}{cc}1&0\\1&1\end{array}\right), \quad L(a)=g_{L}^{a}
\label{Lgenerator}
\end{equation}
and

\begin{equation}
g_{R}=\left(\begin{array}{cc}1&1\\0&1\end{array}\right), \quad R(a)=g_{R}^{a}
\label{Rgenerator}
\end{equation}
respectively.

The "rotation" subgroup $SO_{2}(\mathbb{Z}_N)$:

\begin{equation}
SO_{2}(\mathbb{Z}_N)=\bigg\{ \left( \begin{array}{cc}a&-b\\b&a\end{array} \right), \quad a^2+b^2=1 \bmod N ,\quad a,b\in \mathbb{Z}_N \bigg\}
\end{equation}
By definition $SO_{2}(\mathbb{Z}_N)$ preserves the Euclidean distance $q^2+p^2 \bmod N$. If $N=p^n$ with $p$ prime integer the order $g$ of $SO_{2}[\mathbb{Z}_{p^n}]$ is \cite{Athanasiu:1996}

\begin{equation}
g=\bigg\{ \begin{array}{cc} p^{n-1}(p+1), & p=3 \bmod 4 \\ p^{n-1}(p-1), & p=1 \bmod 4 \end{array}
\end{equation}
They are cyclic for every prime $p\neq2$ and $n$ and therefore they posseses genereator i.e. element $\left( \begin{array}{cc}a_{0}&-b_{0}\\b_{0}&a_{0}\end{array} \right)$, $a_{0},b_{0} \in \mathbb{Z}_{N}$ whose powers generate $SO_{2}(\mathbb{Z}_{p^n})$. It is easy to check from the above the "Fourier" group element i.e. the rotation by $90^{0}$,

\begin{equation}
\epsilon=\left( \begin{array}{cc} 0 & -1 \\ 1 & 0   \end{array} \right)
\label{epsilonDefinition}
\end{equation}
belongs to $SO_{2}(\mathbb{Z}_{p^n})$ and it has order 4. So there is a power, $m$, of $g_{0}$ such that 

\begin{equation}
g_{0}^{m} = \epsilon \bmod p_{n}
\label{g0tom}
\end{equation}
Indeed in the case $p=4k+1$ the order of $SO_{2}(\mathbb{Z}_{p^n})$, $m=p^{n-1}(p-1)=(4k+1)^{n-1} \cdot 4k$ is divisible by 4, so 

\begin{equation}
g_{0}^{ k(4k+1)^{n-1} } = \epsilon 
\end{equation}
and in the case $p=4k+3$ the order $m=p^{n-1}(p+1)=(4k+3)^{n-1} \cdot 4(k+1)$

\begin{equation}
g_{0}^{ (k+1)(4k+3)^{n-1} } = \epsilon 
\label{order4k3}
\end{equation}
We call the generator $g_{0}$ the Fractional Fourier group element since it is the $m$-th root of the Fourier group element. All the powers of $g_{0}$  are the possible fractional powers of the Fourier groups elements and can be seen as fractional rotation on the discrete phase space $\mathbb{T}_{p^n}^{2}$. We notice finally that for $p=1 \bmod 4$ it is possible to determine analytically the generator, while  for $p=3 \bmod 4$ only with trial and error \cite{Athanasiu:1994}.

There is another subgroup of $SL_{2}(\mathbb{Z}_{N})$ which will play an important role in our discussion. This is the scaling or dilation subgroup $D$. For every $a \in \mathbb{Z}_{N}$ which is relative prime to $N$, that is the greatest common divisor  of $a$ and $N$ is equal to $1$ ($GCD(a,N)=(a,N)=1$) we define 

\begin{equation}
D=\Bigg\{D(a)=\left( \begin{array}{cc} a & 0 \\ 0 & a^{-1}   \end{array} \right) \Bigg| \quad a \in \mathbb{Z}_{N}, \quad (a,N)=1   \Bigg\}
\label{Dsubgroup}
\end{equation}
The inverse $a^{-1}$ is defined $\bmod N$ i.e. it is the element of $\mathbb{Z}_{N}$, which satisfies the equation 

\begin{equation}
a^{-1}\cdot a  = 1 \bmod N
\end{equation}
From the above definitions it is easy to see that any element $A\in SL_{2}(\mathbb{Z}_{N})$ can be decomposed as the product (see Eqns. \eqref{Rsubgroup},\eqref{epsilonDefinition},\eqref{Dsubgroup})

\begin{equation}
A=R(x)\cdot D(y) \cdot \epsilon \cdot R(z)
\end{equation}
Indeed, if $A=\left( \begin{array}{cc} a & b \\ c & d   \end{array} \right)$ with $(c,N)=1$  then
\begin{equation}
x=ac^{-1}, y=c^{-1}, z=dc^{-1}
\end{equation}
If $c$ is not coprime with $N$ then $d$ must be coprime with $N$. This follows from the relation $ad-cb=1 \bmod N$. So in this case we decompose the matrix

\begin{equation}
A\epsilon=\left( \begin{array}{cc} b & -a \\ d & -c   \end{array} \right)
\label{AepsilonDecomposition}
\end{equation}
and we get a decomposition of $A$ by inverting Eq. \eqref{AepsilonDecomposition}

\begin{equation}
A=\left( \begin{array}{cc} -b & a \\ -d & c   \end{array} \right)\epsilon
\label{ADecomposition}
\end{equation}
In this paper we shall focus on the elements  $g$ of the "Fractional of Fourier transform group" $SO_{2}(\mathbb{Z}_{N})$

\begin{equation}
g=\left( \begin{array}{cc} a & -b \\ b & a   \end{array} \right)
\label{FrFTgroup}
\end{equation}
which for $b$ coprime with $N$ we get the decomposition 

\begin{equation}
\left( \begin{array}{cc} a & -b \\ b & a   \end{array} \right) = \\
\left( \begin{array}{cc} 1 & ab^{-1} \\ 0 & 1   \end{array} \right)  \\
\left( \begin{array}{cc} b^{-1} & 0 \\ 0 & b   \end{array} \right)  \\
\left( \begin{array}{cc} 0 & -1 \\ 1 & 0   \end{array} \right)  \\
\left( \begin{array}{cc} 1 & ab^{-1} \\ 0 & 1   \end{array} \right) \\
\label{FrFTgroupDecomposition}
\end{equation}
If $b$ is not coprime with $N$ then we follow Eq. \eqref{ADecomposition}.

In the next Section we shall present the construction $g$ of the unitary matrix corresponding to the generator 
\begin{equation}
g_{0}=\left( \begin{array}{cc} a_{0} & -b_{0} \\ b_{0} & a_{0}   \end{array} \right)
\label{g0}
\end{equation}
of $SO_{2}(\mathbb{Z}_{N})$ applying methods of Finite Quantum Mechanics establishing the connection with the quantum mechanical motion of a point particle moving on a discrete circle, $\{S_{N}=1,\omega,\omega ^{2},\ldots , \omega ^{N-1}\}$ with $\omega=e^{\frac{2\pi i}{N}}$.
In later Sections we shall construct quantum circuits which will implement the unitary matrices $U(R(x))$, $U(D(y))$ and $U(\epsilon)=F$ which is the Quantum Fourier Transform (QFT). Finally, we shall construct $U(g_{0})$ which is the quantum circuit for Arithmetic Fractional Fourier Transform.

\section{Finite quantum mechanics and the fractional Fourier transform}\label{sec:FQMandFrFT}

In Quantum Mechanics the point particles of Classical Mechanics are replaced by particle-wave objects which are described by wave functions $\ket{\psi}$, elements of the Hilbert space $\mathcal{H}$ of states and the classical physical observables are replaced by operators - infinite dimensional matrices acting on the Hilbert space  $\mathcal{H}$. For an introduction  to Quantum Mechanics with applications see \cite{Schwinger:2013}.

In FQM the wave particle moves on the set of $N$-discrete points of the unit circle

\begin{equation}
S_{N}=\{1,\omega, \omega ^2 ,\dots , \omega ^{N-1}   \}
\end{equation}
where $\omega= e^{2\pi i/N}$.

The position operator is replaced by the diagonal matrix 

\begin{equation}
Q= \left( \begin{array}{ccccc}
1 & 0 & 0 & \cdots & 0 \\
0 & \omega & 0 & \cdots & 0 \\
0 & 0 & \omega ^{2} & \cdots & 0 \\
\vdots & \vdots & \vdots & \ddots & \vdots \\
0 & 0 & 0 & \cdots & \omega ^{N-1} \\
\end{array} \right)
\label{Q}
\end{equation}
or

\begin{equation}
Q_{k,l}=\omega ^{k} \delta _{k,l} , \quad k,l=0,1,\ldots,N-1
\end{equation}
The state of a particle-wave sitting in the position $\omega ^{k}, \, k=0,1,\ldots,N$,  is the vector

\begin{equation}
\ket{e_{k}}=\left( \begin{array}{c} 
0 \\ 0 \\ \vdots \\ 1 \\ \vdots \\ 0 \\0
\end{array} \right), \quad  \langle e_{l} | e_{k} \rangle = \delta _{l,k}
\label{ek}
\end{equation}
deriving  $Q\ket{e_{k}}=\omega ^{k}\ket{e_{k}}, \, k=0,\ldots , N-1$.

The translation-momentum operator is represented by the matrix 

\begin{equation}
P= \left( \begin{array}{ccccc}
0 & 0 & 0 & \cdots & 1 \\
1 & 0 & 0 & \cdots & 0 \\
0 & 1 & 0 & \cdots & 0 \\
\vdots & \vdots & \vdots & \ddots & \vdots \\
0 & 0 & 0 & \cdots & 0 \\
\end{array} \right)
\end{equation}
or

\begin{equation}
P_{k,l}= \delta _{k-1,l} , \quad k,l=0,1,\ldots,N-1
\end{equation}
It holds that $P\ket{e_{k}}=\ket{e_{k+1}},\, k,l=0,\ldots , N-1$, so it moves the partile from position $k$ to $k+1$.

The eigenstates of $P$ are the plane waves 

\begin{equation}
\ket{p_{k}}=\frac{1}{\sqrt{N}}\left( \begin{array}{c} 
1 \\ \omega ^{k} \\ \omega ^{2k}  \\ \vdots \\ \\\omega ^{(N-1)k}
\end{array} \right), \quad  k=0,1,\ldots, N-1
\end{equation}
so it holds 

\begin{equation}
P\ket{p_{k}}=\omega ^{-k} \ket{p_{k}} , \quad k,=0,1,\ldots,N-1
\label{Pk}
\end{equation}

Eqns. \eqref{Q},\eqref{ek} and \eqref{Pk} imply that the diagonalizing matrix of $P$ is $F$ :

\begin{equation}
F_{k,l}=\frac{1}{\sqrt{N}}\omega ^{kl},\quad  k,l=0,1,\ldots, N-1
\end{equation}
so

\begin{equation}
QF=FP
\label{QF}
\end{equation}
or 

\begin{equation}
PF^{\dagger}=F^{\dagger}Q
\label{PFdagger}
\end{equation}
which is the QFT matrix. Eqns. \eqref{QF} and \eqref{PFdagger} express the particle-wave dualtiy nature of quantum mechanical point particles. Notice that $Q^{N}=P^{N}=I_{N\times N}$ (periodic boundary condition for wave function around the torus $\mathbb{T}_{N}^{2}$).

The classical discrete phase-space of a point particle consists of points 

\begin{equation}
(r,s) \in \mathbb{Z} _{N} \times \mathbb{Z} _{N} = \mathbb{T} _{N}^{2}, \quad \mathbb{Z}_{N} = \{ 0,1,2,\ldots , N-1 \}
\end{equation}
In FQM to every point $(r,s) \in  \mathbb{T} _{N}^{2}$ we assign the translation matrix, from the point $(0,0)$ to $(r,s)$, 

\begin{equation}
J_{r,s} =\omega ^{\frac{rs}{2}}P^{r}Q^{s}
\end{equation}
where the $\frac{1}{2}$ in the exponent for $N$ odd is equal to $\frac{N+1}{2} \bmod N$. If $N$ is even we must define $J_{r,s}$ in a different way \cite{Zak:1989}.

The basic matices $Q$,$P$, called also clock and shift matrices, satisfy the "exponential" Heisenberg uncertainty relation

\begin{equation}
QP=\omega PQ
\label{QP}
\end{equation}
This exponential relation has been compared \cite{Floratos:1997} to the standard Heisenberg canonical commutation relation of QM

\begin{equation}
\left[ \hat{q}, \hat{p} \right] = i\hbar
\end{equation}
where $\hat{q}$ and $\hat{p}$ are the position and momentum operators.

Using Eq. \eqref{QP} we can check the following important properties of $J_{r,s}$ which are called in the physics literature "magnetic relations"

\begin{equation}
J_{r,s}J_{r^{\prime},s^{\prime}}= \omega ^{\frac{1}{2} \tiny{\begin{pmatrix} r, s  \end{pmatrix}} \epsilon  \tiny{\begin{pmatrix} r^{\prime} \\ s^{\prime} \end{pmatrix}}  } J_{r+r^{\prime},s+s^{\prime}}
\label{JrsJrs}
\end{equation}

\begin{equation}
J_{r,s}^{k} = J_{kr,ks}
\end{equation}

\begin{equation}
J_{r,s}^{\dagger}=J_{-r,-s}
\end{equation}

\begin{equation}
J_{r,s}^{N} = I_{N\times N}
\end{equation}
The phase in Eq. \eqref{JrsJrs} is equal to the area of the triangle in $\mathbb{T}_{N}^{2}$ with vertices $(0,0)$, $(r,s)$, $(r^{\prime},s^{\prime})$. Eq. \eqref{JrsJrs} implies that the motion of a particle around the parallelogram is executed by the operators 
\begin{equation}
J_{r,s}J_{r^{\prime},s^{\prime}} J_{r,s}^{\dagger}J_{r^{\prime},s^{\prime}}^{\dagger}=\omega ^{ \tiny{\begin{pmatrix} r,s \end{pmatrix}} \epsilon  \tiny{\begin{pmatrix} r^{\prime} \\ s^{\prime} \end{pmatrix}}  }
\end{equation}
%\omega ^{(r^{\prime},s^{\prime}) \epsilon \left( \begin{array}{c} r \\ s \end{array} \right)}
This is known in the physics literature as the Aharanov-Bohm effect \cite{Aharonov:1959}, for electron moving around a magnetic solenoid.

After the preliminary discussion we shall continue with the basic steps of the construction of a particular repesentation of the LCT group $SL_{2}(\mathbb{Z}_N)$, called the Weil metaplectic representation \cite{Feichtinger:2007}.
We consider only the case of odd $N$ due to number theoretic intricacies of the construction for the case of even $N$ \cite{Feichtinger:2007}.

To every element $A \in SL_{2}(\mathbb{Z}_N)$ we assign a $N$-dimensional unitary matrix $U(A)$ with the constraint that 

\begin{equation}
U^{\dagger}(A)J_{r,s}U(A)=J_{(r,s)A}
\label{UAconstraint}
\end{equation}
In physics terms the one-time step motion of a particle in the phase-space  $\mathbb{T}_{N}^{2}$  described in classical mechanics by $(r,s)\rightarrow (r,s)A \bmod N$, \, $A \in SL_{2}(\mathbb{Z}_N)$, is expressed in FQM as one-time step evolution of the wave-fucntion $\ket{\psi}$ of a particle  by $\ket{\psi}\rightarrow U(A)\ket{\psi}$ and this transforms the matrices $J_{r,s}$ as $J_{r,s}\rightarrow U^{\dagger}(A)J_{r,s}U(A)$, which is the left hand side of Eq. \eqref{UAconstraint}.

The constraint \eqref{UAconstraint} determines $U(A)$ up to an overall (global) phase. Moreover this relation implies that:

\begin{equation}
U(A)U(B)=\omega ^{\varphi (A,B)}  U(AB)
\end{equation}
From the associativity of the matrices 

\begin{equation}
(U(A)U(B))U(C)=U(A)(U(B)U(C))
\end{equation}
we get a constraint for $\varphi (A,B)$

\begin{equation}
\varphi (A,B) + \varphi (AB,C) = \varphi (A,BC) + \varphi (B,C)
\label{phiAB}
\end{equation}
This relation defines the phase $\varphi$ as a 2-cocycle of the group $SL_{2}(\mathbb{Z}_N)$. Possible solutions of Eq. \eqref{phiAB} give different representation for $SL_{2}(\mathbb{Z}_N)$ \cite{Feichtinger:2007}.

Explicit forms of the unitary matrices of the Weil representation in the case of $N=p$ or $N=p^{n}$, where $p$ is a prime, have been presented in \cite{Athanasiu:1994}.  We follow \cite{Athanasiu:1996} where detailed derivations are presented and for particular forms of $U(A)$  with phase $\varphi(A,B)=0$.

Up to an overall phase the matrix elements of $U(A)$ are given as follows: 

\noindent
For generic $A\in SL_{2}(\mathbb{Z}_N)$

\begin{equation}
U(A)_{k,l}=\frac{1}{\sqrt{N}} \omega ^{-\frac{ak^2-2kl+dl^2}{2c}}, \quad k,l=0,1,\ldots, N-1
\label{UA}
\end{equation}
For $A=\left(\begin{array}{cc}a&0\\0&a^{-1}\end{array}\right), \quad (a|N)=1, \quad a\in \mathbb{Z}_N$

\begin{equation}
U(A)_{k,l}=\delta _{ak,l}, \quad k,l=0,1,\ldots, N-1
\label{Umul}
\end{equation}
For $A=\left(\begin{array}{cc}1&a\\0&1\end{array}\right), \quad (a|N)=1, \quad a\in \mathbb{Z}_N$

\begin{equation}
U(A)_{k,l}=\omega ^{-\frac{1}{2}ak^{2}}, \quad k=0,1,\ldots, N-1
\label{Udiag}
\end{equation}
We notice  the important fact that $\epsilon = \left(\begin{array}{cc}0&-1\\1&0\end{array}\right) \in SL_{2}(\mathbb{Z}_N)$ is represented by the QFT matrix (up to a global phase)

\begin{equation}
U(\epsilon)_{k,l} \sim F_{k,l}= \frac{1}{\sqrt{N}} \omega ^{kl}, \quad k,l=0,1,\ldots,N-1
\label{Uepsilon}
\end{equation}
At this point we are ready to present our proposal for the Fractional Fourier Transform. From Section \ref{sec:LCT}, Eqns. \eqref{g0} \eqref{UA} we get for the generator of the discrete rotation group $g_{0}$

\begin{equation}
U( \left(\begin{array}{cc}a_{0}&-b_{0}\\b_{0}&a_{0}\end{array}\right) )_{k,l} \sim \frac{1}{\sqrt{N}} \omega ^{-\frac{a_{0}(k^{2}+l^{2})-2kl}{2b_{0}}}
\label{AFrFT}
\end{equation}
We observe the power on Eq. \eqref{g0tom} connecting $g_{0}$ and $\epsilon$ in $SL_{2}(\mathbb{Z}_N)$ ($g_{0}$ is the $m$-th root of $\epsilon$)

\begin{equation}
g_{0}^{m}=\epsilon
\end{equation}
From Eq. \eqref{g0tom} we get

\begin{equation}
U(g_{0}^{m})=U(\epsilon)\sim F \implies U(g_{0})^{m}=F
\end{equation}
So indeed $U(g_{0})$ is the $N\times N$ unitary matrix which is the $m$-th root ($1/m$) of the Fourier matrix $F$.

The representation $U(g_{0})$,  generates all the fractional powers of the QFT matrix $F$.

\begin{equation}
U(g_{0}), U^{2}(g_{0}),\ldots, U^{m}(g_{0})=F, \ldots , U^{4m}(g_{0})=I_{N\times N}
\end{equation}
In the following Sections we shall construct the quantum circuits for $U(g_{0})$ using the decomposition of Eq. \eqref{FrFTgroupDecomposition} from Section \ref{sec:LCT}.

\section{Multilevel qudit circuits and the QFT}\label{sec:QFTQudits}

The states of a $p$-level qudit span a complex vector space of $p$ dimensions whose computational basis states are denoted $\ket{0},\ket{1},\ldots ,\ket{p-1}$. A collection of $n$ qudits spans the $n$-fold tensor product of such $p$-dimensional spaces and this tensor product has dimension $p^{n}$. The computational basis of such a collection is the set $\{\ket{b}:b=0\ldots p^{n}-1 \}$. Similarly to the two dimensional qubits case, we can use the $p$-ary representation of an integer $b$ which is $b=\left( b_{n-1}b_{n-2}\ldots b_{0} \right) =\sum_{j=0}^{n-1}b_{j}p^{j}$ with $b_{j}=0\ldots p-1$. We shall use the name "dits" for the $p$-ary digits of such a representation. Moreover, the fractional $p$-ary notation $(0.b_{n-1}\ldots b_{1}b_{0})=\sum_{j=0}^{n-1}b_{j}p^{j-n}=b/p^{n}$ will be used. Using these notations, any computational basis state on $n$ qudits can be expressed as the tensor product $\ket{b}=\ket{b_{n-1}}\ket{b_{n-2}}\cdots \ket{b_{0}}$. In this manuscript we restrict the dimensionality $p$ to be a prime number as demanded by the AFrFT application.

While the motivation to design multilevel qudit circuits is the quantum implementation of the AFrFT developed in the previous sections, the proposed circuits have a broader appllication alone, e.g. multipliers are integrant part of quantum phase estimation (integer factoring, discrete logarithm problem \cite{Shor:1994} etc.) and quadratic phases diagonal operators find application in quantum system simulations \cite{Nielsen:2011}. Some advantages of multilevel qudits over two-level qubits are related to the decoherence problem and at a higher level to performance improvement of specific quantum algorithms. A brief review of such advantages along with references is given in \cite{Pavlidis:2021}.

\subsection{Qudit gates}\label{subsec:QuditGates}

Qudit gates are unitary transformations of states of qudits collection. It is proven that single and two-qudit gates are universal \cite{Brylinski:2002}, so they are enough to construct any qudit circuit. On systems of $p$-level qudits the single qudit and two-qudit  gates are represented by unitary matrices of dimensions $p\times p$ and  $p^{2}\times p^{2}$, respectively. Usually an \emph{elementary} single qudit gate is defined to operate on a two-dimensional subspace (two-level gate) of the whole $d$-dimensional space. Similarly, two-qudit elementary gates are usually defined to operate on a two-dimensional subspace of the target qudit conditioned on a particular state of the control qudit. A variety of such elementary gates can be found in the literature \cite{Brylinski:2002,Muthukrishnan:2000,Di:2013}. Based on such elementary gates, more complex single qudit gates can be constructed which operate on the whole $p$-dimensional qudit space or on the whole $p^{2}$-dimensional qudit space for the two-qudit case. 

The gates which will be used extensively in this and the subsequent Sections are generalizations  on $p$-dimensional qudits of the qubit Hadamard gate, single qubit $z$-axis rotation gates $R_{z}$ and two-qubit controlled  $z$-axis rotation gates $R_{z}$. These generalizations are indispensable for the construction of the QFT on $n$ qudits \cite{Muthukrishnan:2002,Ermilov:2007} as well as  for our proposed constructions. We shall use an "exponent" notation $^{(p)}$ in the qudit gates symbols and the more complex utilized operators to emphasize that they are defined in the context of $p$-dimensional qudits quantum circuits. 

The single $p$-level qudit Hadamard  gate is represented by the operator

\begin{equation}\label{eq:Hd}
H^{(p)}=  \frac{1}{\sqrt{p}} \sum_{j=0}^{p-1}  \sum_{k=0}^{p-1} \ket{j} \bra{k} e^{i2\pi \frac{jk}{p}}
\end{equation}

\noindent
It is worth to consider the effect of a Hadamard gate on a basis state $\ket{b}$ :

\begin{equation}\label{eq:Hd2basis}
H^{(p)}|b \rangle  = \frac{1}{\sqrt{p}} (   \ket{0}+e^{i2\pi(0.b)}\ket{1}+\cdots +e^{i2\pi(p-1)(0.b)} \ket{p-1}   )
\end{equation}

\noindent
where $(0.b)$ is the fractional $p$-ary representation of $b/p$. Physical implementations of $p$-level Hadamard qudit gate are proposed in \cite{Muthukrishnan:2002,Ermilov:2007}.

The single qudit rotation gate corresponding to the $R_{z}(\theta)$ qubit gate is the diagonal gate defined by $R_{z}^{(p)}(\theta _{1},\ldots , \theta _{p-1})=\diag(1,e^{i\theta _{1}},\ldots , e^{i\theta _{p-1}})$.  Two-qudit extension of this diagonal gate is defined with $R_{z}^{(p)}(\theta _{1},\ldots , \theta _{p^{2}-1})=\diag(1,e^{i\theta _{1}},\ldots , e^{i\theta _{p^{2}-1}})$. Synthesis of such gates using elementary two-level single qudit and two-qudit gates can be found in \cite{Di:2013,Pavlidis:2021}. The same symbol $R^{(d)}$ will be used for both the single and two-qudit gates; the differentiation will be clear by the context or by an extra vertical line controlling the target qudit in the diagrams, as is the usual practice on qubits circuits. We shall use specific forms of these gates,  where all the arbitrary angles $\theta_{l}$ in each gate are determined by a single parameter $k$. Namely, the controlled two-qudit rotation gate $R_{k}^{(p)}$ is defined by the relation

\begin{equation}\label{eq:Rkfull}
R_{k}^{(p)}=
\sum_{j=0}^{p-1} \sum_{m=0}^{p-1} 
e^{i\frac{2\pi}{p^{k}}jm}
\ket{j} \bra{j} \otimes \ket{m} \bra{m} 
\end{equation}
With the above definition the arbitrary angles $\theta _{l}$ for $l=jp+m$ and $j,m=0\ldots p-1$ take the specific values $(2\pi/p^{k})jm$, in analogy with the qubit controlled rotation gates. The gate $R_{k}^{(d)}$ changes the superposition phases of a target qudit, depending on the state of a control qudit. In particular, assuming that the control qudit is in a basis state $\ket{b}$ and the target qudit is in a superposition state $\ket{s}=\frac{1}{\sqrt{p}}\sum_{l=0}^{p-1}e^{i\varphi_{l}}\ket{l}$, then the effect of the  $R_{k}^{(p)}$ gate on both qudits is

\begin{equation}\label{eq:Rkeffect}
R_{k}^{(p)} \left( \ket{b} \ket{s} \right) = \frac{1}{\sqrt{p}} 
\ket{b} \sum_{m=0}^{p-1} 
e^{i \tiny{\left( 2\pi (0.{\underbrace{00\ldots 0}_{k-1}}b)m +\varphi_{m} \right)} }
\ket{m}
\end{equation}

\noindent
The single qudit gate $R_{k}^{(d)}$ is defined by the operator

\begin{equation}\label{eq:singleRkfull}
R_{k}^{(p)}=
\sum_{m=0}^{p-1} 
e^{i\frac{2\pi}{p^{k}}m^{2}}
\ket{m} \bra{m} 
\end{equation}

\noindent
The Hadamard gate defined in Eq. \eqref{eq:Hd} and the single and two-qudit gates defined in Eqns. \eqref{eq:Rkfull} and \eqref{eq:singleRkfull} are the basic gates which will be used for the design of the QFT, the in-place multiplier and the diagonal operator required for the the QAFrFT circuit. Quantum cost and depth analysis of the single and two qudit $R_{k}^{(p)}$ gates in terms of elementary two-level qudit gates can be found in \cite{Pavlidis:2021}. 

\subsection{QFT on qudits}\label{subsec:QuditQFT}

The QFT on $n$ qudits of $p$ levels is defined on the computational basis by the transformation 

\begin{equation}\label{eq:QFTqudits}
\begin{gathered}
\ket{j}=\ket{j_{n-1}j_{n-2}\ldots j_{0}} \xrightarrow{\text{QFT}_{p^{n}}} 
\frac{1}{\sqrt{p^{n}}}
\sum_{k=0}^{p^{n}-1} 
e^{\frac{i2\pi}{p^{n}}jk}\ket{k} =  \\
\frac{1}{\sqrt{p^{n}}}\left( 
\sum_{m=0}^{p-1}
e^{i2\pi (0.j_{0}) m} \ket{m}
\right)
\left( 
\sum_{m=0}^{p-1}
e^{i2\pi (0.j_{1}j_{0}) m} \ket{m}
\right)
\cdots \\
\cdots \left( 
\sum_{m=0}^{p-1}
e^{i2\pi (0.j_{n-1}j_{n-2}\ldots j_{1}j_{0}) m} \ket{m}
\right) 
\end{gathered}
\end{equation}

\begin{figure}[!h]
	\centering{\epsfig{file=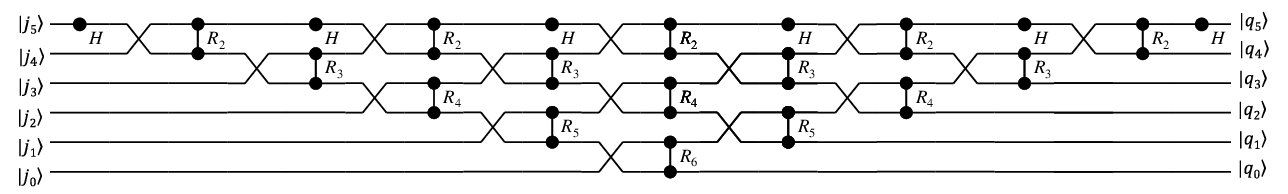, width=14cm}}
	\vspace*{13pt}
	\caption{\label{fig:QFTQuditsLocal} QFT circuit with local interactions on 6 qudits. Single qudit Hadamard gates $H^{(p)}$ are denoted with a bullet, while two qudit rotation gates $R_{k}^{(p)}$ are denoted with two bullets connected with a vertical line. Swap gates are denoted with crossing lines. Note that the qudits are rearranged in the correct reverse order at the end of the computation.}
\end{figure}

\begin{figure}[!h]
	\centering{\epsfig{file=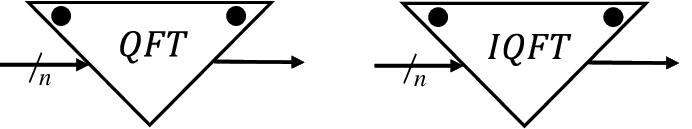, width=8cm}}
	\vspace*{13pt}
	\caption{\label{fig:QFTQuditsLocalSymbol} Symbols for QFT and IQFT circuits with local interactions Thick bullets at the input and ouput sides denote the most significant qudit positions.}
\end{figure}

The above  product state form derived  permits to construct the QFT circuit on $p$-level qudits like the two-level qubits case. Namely, the topology of the QFT on $n$ qudits of $p$ levels is exactly the same with the one for the qubits case, replacing qubit Hadamard gates  with the qudits Hadamard gates $H^{(p)}$ defined in Eq. \eqref{eq:Hd} and replacing qubits controlled rotation gates $R_{k}$ for $k=2\ldots n$, with the qudits controlled rotation gates $R_{k}^{(p)}$ defined in Eq. \eqref{eq:Rkfull}. Detailed analysis of such a construction is given in \cite{Muthukrishnan:2002,Ermilov:2007,Pavlidis:2021}. An inverse QFT (IQFT) circuit can be constructed by fliping left-right the QFT circuit and applying opposite angles in the rotation gates.
Due to the identical topology with the qubits QFT, it can be mapped on an one-dimensional local nearest neighborhood (1D-LNN) architecture which permits local interactions only, e.g. \cite{Fowler:2004}, using qudits swap gates \cite{Escartin:2013}. For the sake of completeness such a configuration is shown in Figure \ref{fig:QFTQuditsLocal} as it will be used later. The transformed joint state appears in the correct reverse qudits order at the end, thus there is not need for extra swap gates. This topology offers linear depth  provided that the architecture is capable of parallel execution of gates.  The depth of this circuit is $2n-1$ steps counting adjacent gates acting on same qudits as one step (e.g. SWAP and two qudit controlled gates are considered as merged into one gate), while the quantum cost is $n(n+1)/2$ counting in the same manner. Also, observe that initially most of the qudits remain idle and as we proceed from left to right, more qudits are involved in gates execution until the middle of the circuit where all the qudits are engaged in gates executions. After the middle more and more qudits remain idle. This fact may be exploited by adjacent subcircuits to execute their gates on the idle qubits, leading to a pipeline processing. For this reason we adopt the triangular symbols shown in Figure \ref{fig:QFTQuditsLocalSymbol} for the QFT and IQFT circuits.

\section{QFT-based in-place constant Multiplier on \\ multilevel qudits}\label{sec:MultQudits}
%\subsection{QFT-based Multiplier modulo $p^{n}$ by constant $\lambda$ co-prime with $p$}\label{subsec:MultOddModQudits}
In this Section we develop the multiplier required in the AFrFT quantum circuit, which is a modulo  multiplier by constant operating on $p$-level qudits. The need for such a quantum circuit arises from the representation \eqref{Umul} and the decomposition \eqref{FrFTgroupDecomposition}. Indeed, the action of the unitary matrix $MUL_{\lambda}=\delta _{k,\lambda l}$ on the basis state $\ket{l}=\ket{e_{l}}=  \left( \begin{array}{ccccccc} 0 & 0 & \cdots & 1 & \cdots & 0 & 0 \end{array} \right)^{T} $ is

\begin{equation}
MODMULC_{\lambda}\ket{l}=\sum_{l=0}^{p^{n}-1} \delta _{k,\lambda l} \ket{e_{l}} = \ket{e_{\lambda l}}=\ket{\lambda l}
\label{eq:MODMULCDefQudits}
\end{equation}
where the multiplication is done $\bmod p^{n}$.

Our proposal for the quantum modular multiplication circuit is based in the QFT representation of integers instead of the usual computational base representation. It is an "in-place" multiplier in the sense that it doesn't use any ancilla qudits. This multiplier  by constant $\lambda$  performs the multiplication modulo $p^n$, where $n$ is the qudits number (each one of $p$ dimensions). Variations and extensions of this circuit are given in Section \ref{sec:Variations}. Unitarity of the multiplier poses the restriction $\gcd(\lambda,p^{n}) = 1$. The proposed design relies on the definition of a new operator which we call \textit{modified QFT}  with parameter $\lambda$, denoted from now on with $mQFT_{\lambda}^{(p)}$. This operator on $n$ qudits is defined by

\begin{equation}\label{eq:mQFTopQudits}
mQFT_{\lambda}^{(p)}=\frac{1}{\sqrt{p^{n}}} \sum_{k=0}^{p^{n}-1} \sum_{j=0}^{p^{n}-1} \ket{k}\bra{j} e^{i\frac{2\pi}{p^{n}}\lambda jk}
\end{equation}

\noindent
The above definition resembles that of the usual QFT in Eq. \eqref{eq:QFTqudits} justifying its name. The motivation for this definition is the fact that a succesive application of an $mQFT_{\lambda}^{(p)}$ operator and an inverse QFT  results in the multiplier by constant $\lambda$. Now we prove this claim:

\begin{equation}
\begin{split}
MODMULC_{\lambda}^{p}=  QFT^{(p)-1}\cdot mQFT_{\lambda}^{(p)} = \\
\frac{1}{p^{n}}   \sum_{m=0}^{p^{n}-1} \sum_{r=0}^{p^{n}-1} \ket{m} \bra{r} e^{-i\frac{2\pi}{p^{n}}mr}  \sum_{k=0}^{p^{n}-1} \sum_{j=0}^{p^{n}-1} \ket{k}\bra{j} e^{i\frac{2\pi}{p^{n}}\lambda jk }=  \\
%% & \frac{1}{p^{n}}   \sum_{m=0}^{p^{n}-1} \sum_{r=0}^{p^{n}-1}  \sum_{k=0}^{p^{n}-1} \sum_{j=0}^{p^{n}-1}  \ket{m} \delta _{rk} \bra{j} e^{i\frac{2\pi}{p^{n}} (\lambda jk - mr)} = \\
 \frac{1}{p^{n}}   \sum_{m=0}^{p^{n}-1}   \sum_{j=0}^{p^{n}-1}  \ket{m}  \bra{j}  \sum_{k=0}^{p^{n}-1} \left( e^{i\frac{2\pi}{p^{n}} (\lambda j - m)} \right)^{k} = \\
     \sum_{j=0}^{p^{n}-1}  \ket{\lambda j \bmod p^{n}}  \bra{j}  
\end{split}
\label{eq:IQFTmQFTopQudits}
\end{equation}

\noindent
The next to last equation holds due to the fact that the sum of roots of unity is zero, explicitly

\begin{equation}\label{eq:rootsd}
\sum_{k=0}^{p^{n}-1} \left( e^{i\frac{2\pi}{p^{n}} (\lambda j - m)} \right)^{k} = 
\begin{cases}
p^{n},	&  \lambda j -m = 0 \pmod{p^{n}} \\
0, 		&   \lambda j -m \neq 0 \pmod{p^{n}}
\end{cases}
\end{equation}

\noindent
Clearly, the operator derived in Eq. \eqref{eq:IQFTmQFTopQudits} transforms any computational base state $\ket{l}$ to $\ket{\lambda l \bmod p^n}$, which is exactly the definition of Eq. \eqref{eq:MODMULCDefQudits} multiplier.

It remains to give an efficient quantum circuit construction for the $mQFT_{\lambda}^{(p)}$ operator. The resemblence between Eq \eqref{eq:mQFTopQudits} and Eq. \eqref{eq:QFTqudits} lead us to the circuit of Fig. \ref{fig:mQFTQudits},  which has almost identical topology with the usual QFT circuit, except two key distinctions: 

\begin{enumerate}[(a)]
	\item Each controlled rotation gate $R_{k}^{(p)}$ of the QFT is replaced with the controlled rotation gate $R_{k}^{(p)\lambda}$. This is equivalent to say that the angles used in the phases of these new rotation gates are multiples by $\lambda$  of the original angles used in the rotation gates of the usual QFT.
	\item A single qudit permutation gate $P_{\mu}^{(p)}$ is applied after each Hadamard gate. This permutation gate performs the operation
	
	\begin{equation}\label{eq:PQudits}
	\ket{x} \xrightarrow{P_{\mu}^{(p)}} \ket{\mu x \bmod p}
	\end{equation}
	
	\noindent 
	for $\mu =\lambda ^{-1} \pmod{p}$. Such a gate can be easily constructed by decomposing the desired permutation into a product of transpositions which correspond to the single qudit elementary gates $X^{(jk)}=\ket{j} \bra{k} + \ket{k} \bra{j} +\sum_{m\neq j,k}\ket{m} \bra{m}$  operating on a two-level subspace \cite{Di:2013}. The length of this sequence is at most $p-1$. 
\end{enumerate}

\renewcommand{\rgate}[1]{*=<2.5em>[Fo]{#1} \qw} 
\begin{figure}[!h]
	\scalebox{0.70}{
	\Qcircuit @C=0.5em @R=0.35em {
				\lstick{\ket{j_{n-1}}}   &   \gate{H^{(p)}}  & \gate{P_{\lambda ^{-1}}^{(p)}} &  \rgate{R_{2}^{(p)\lambda}} & \qw & \cdots & &   \rgate{R_{n-1}^{(p)\lambda}} &   \rgate{R_{n}^{(p)\lambda}}   &   \qw        &   \qw           &   \qw      &   \qw           &   \qw        &   \qw     &   \qw           &   \qw &   \qw   &\qw & \qw &\qw &\qw &\qw  &  \qw & \rstick{\ket{q_{n-1}}} \qw    & \push{\rule{5em}{0em}}   \\
				\lstick{\ket{j_{n-2}}}   &   \qw       &  \qw & \ctrl{-1}     & \qw  &  \qw  &   \qw   & \qw   &   \qw           &   \gate{H^{(p)}} & \gate{P_{\lambda ^{-1}}^{(p)}} &\qw & \cdots & &    \rgate{R_{n-2}^{(p)\lambda}}         &   \rgate{R_{n-1}^{(p)\lambda}} &\qw & \qw  & \qw &   \qw    &\qw    &   \qw     &   \qw                   & \qw&  \rstick{\ket{q_{n-2}}} \qw       \\
				\lstick{\vdots }         &             &      &           & &        \ddots         &      &                 &              &                 &  &   &      \ddots           &               &    &           &&&      & & & &            & &  \rstick{\vdots }             \\
				\lstick{\ket{j_{1}}}   &   \qw       &  \qw & \qw   & \qw & \qw   & \qw     &   \ctrl{-3}      &   \qw      &  \qw       &\qw &\qw  & \qw      &   \qw        &   \ctrl{-2}     &   \qw       &\qw        &  \cdots & & \gate{H^{(p)}}  & \gate{P_{\lambda ^{-1}}^{(p)}}    &   \rgate{R_{2}^{(p)\lambda}}   &   \qw        & \qw &  \rstick{\ket{q_{1}}} \qw      \\
				\lstick{\ket{j_{0}}}     &   \qw       &   \qw & \qw           &   \qw           &   \qw           &    \qw       &   \qw   &   \ctrl{-4}        &   \qw  & \qw &  \qw &   \qw &   \qw &   \qw    &   \ctrl{-3}  &\qw & \cdots  & &  \qw      &\qw       &   \ctrl{-1}     &   \gate{H^{(p)}}   &  \gate{P_{\lambda ^{-1}}^{(p)}}& \rstick{\ket{q_{0}}} \qw
			}
	
	}
	\vspace*{13pt}
	\caption{\label{fig:mQFTQudits} Modified Quantum Fourier Transform circuit on $n$ qudits of $p$-levels with parameter $\lambda$ where $\gcd(\lambda,p)=1$ (note that the order of the qubits must be reversed at the end of the computation). In the above notation it is assumed that the computational basis input state $\ket{j}$ is expressed with $\ket{j_{n-1}}\ket{j_{n-2}} \cdots \ket{j_{0}}$ in $p$-ary representation. The output states are $\ket{q_{r}}= \frac{1}{\sqrt{p}}\sum_{m=0}^{p-1} e^{i2\pi(0.j_{r}j_{r-1}\ldots j_{0}  )m}\ket{m}$ and they are unentangled. Observe that when  $\lambda=1$ this circuit coincides with the usual $n$ qudits QFT.}
\end{figure}
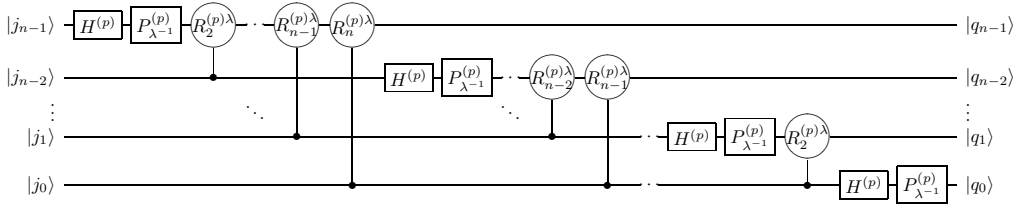 
An arbitrary computational input state $\ket{j}=\ket{j_{n-1}}\cdots \ket{j_{1}}\ket{j_{0}}$ is assumed. We start from the upper qudit (most significant), which is assumed initially in the state $\ket{j_{n-1}}$. At first, gate $H^{(p)}$ applies the transformation  $\ket{j_{n-1}} \xrightarrow{H^{(p)}} \frac{1}{\sqrt{p}} \sum_{m=0}^{p-1} e^{i2\pi(0.j_{n-1} )m}\ket{m}$ (see Eq. \eqref{eq:Hd2basis}). Next, the permutation gate $P_{\lambda ^{-1}}^{(p)}$  evolves this state as

\begin{equation}\label{PeffectQudits}
\begin{split}
\frac{1}{\sqrt{p}} \sum_{m=0}^{p-1} e^{i2\pi(0.j_{n-1}  )m}\ket{m} \xrightarrow{P_{\lambda ^{-1}}^{(p)}} & \frac{1}{\sqrt{p}} \sum_{m=0}^{p-1} e^{i2\pi(0.j_{n-1}  )m}\ket{\lambda ^{-1} m \pmod{p}} = \\
& \frac{1}{\sqrt{p}} \sum_{m^{\prime}=0}^{p-1} e^{i2\pi(0.j_{n-1}  )\lambda m^{\prime}}\ket{m^{\prime} }  
\end{split}
\end{equation}
In the above equation a change of index $m^{\prime}=\lambda ^{-1} m \bmod{p}$ took place. For this remapping is bijective $m^{\prime}=0 \ldots p-1$, the summation limits in the last equation remained the same.
Taking into account Eq. \eqref{eq:Rkeffect}, we find the effect of gate $R_{2}^{(p)\lambda}$ in the state of Eq. \eqref{PeffectQudits} which is $\frac{1}{\sqrt{p}} \sum_{m=0}^{p-1} e^{i2\pi(0.j_{n-1}j_{n-2}  )\lambda m}\ket{m}$. Similar calculations for the effect of all the subsequent rotation gates from $R_{3}^{(p)\lambda}$ to $R_{n}^{(p)\lambda}$ on the upper qudit of Figure \ref{fig:mQFTQudits}   yield the final state $\ket{q_{n-1}}= \frac{1}{\sqrt{p}} \sum_{m=0}^{p-1} e^{i2\pi(0.j_{n-1}j_{n-2}\ldots j_{0}  )\lambda m}\ket{m}$. 

\begin{figure}[!h]
	\centering{\epsfig{file=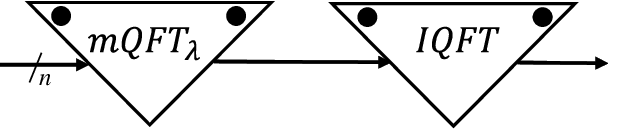, width=8cm}}
	\vspace*{13pt}
	\caption{\label{fig:ModMultOddBlockQudits} The constant multiplier $MODMULC_{\lambda}$ for $\gcd(\lambda ,p^{n})=1$ consisting of a modified QFT with parameter $\lambda$ and an inverse QFT on $n$ qudits, in succession.}
\end{figure}

A similar analysis for the rest of the qudits initally in state $\ket{j_{k}}, k=n-2\ldots 0$ leads to final states 

\begin{equation}
\ket{q_{k}}= \frac{1}{\sqrt{p}} \sum_{m=0}^{p-1} e^{i2\pi(0.j_{k}j_{k-1}\ldots j_{0}  )\lambda m}\ket{m}
\end{equation}
\noindent
Reversing the order of the qudits at the end (something not shown in Fig. \ref{fig:mQFTQudits} like the usual QFT case)  we have the final product state 

\begin{equation}\label{eq:mQFTqudits}
\begin{split}
\ket{q_{0}}\ket{q_{1}}\cdots \ket{q_{n-1}} = 
\frac{1}{\sqrt{p^{n}}}\left( 
\sum_{m=0}^{p-1}
e^{i2\pi \lambda (0.j_{0}) m} \ket{m}
\right)
\left( 
\sum_{m=0}^{p-1}
e^{i2\pi \lambda  (0.j_{1}j_{0}) m} \ket{m}
\right)
\cdots \\
\cdots
\left( 
\sum_{m=0}^{p-1}
e^{i2\pi \lambda  (0.j_{n-1}j_{n-2}\ldots j_{1}j_{0}) m} \ket{m}
\right)  = %\\
\frac{1}{\sqrt{p^{n}}}
\sum_{k=0}^{p^{n}-1} 
e^{\frac{i2\pi}{p^{n}}\lambda jk}\ket{k} 
\end{split}
\end{equation}

\noindent
which indeed corresponds to the  transformation imposed by the $mQFT_{\lambda}^{(p)}$ operator on an arbitrary computational basis state $\ket{j}$. We can see that the above state is the QFT of the state $\ket{\lambda j \bmod p^{n}}$. Consequently,  to recover the multiple $\ket{\lambda j \bmod p^{n}}$ it suffices to apply an inverse QFT after the modified QFT as shown in the block diagram of Fig. \ref{fig:ModMultOddBlockQudits} and we have already proven in Eq. \eqref{eq:IQFTmQFTopQudits}.

\begin{figure}[!h]
	\centering{\epsfig{file=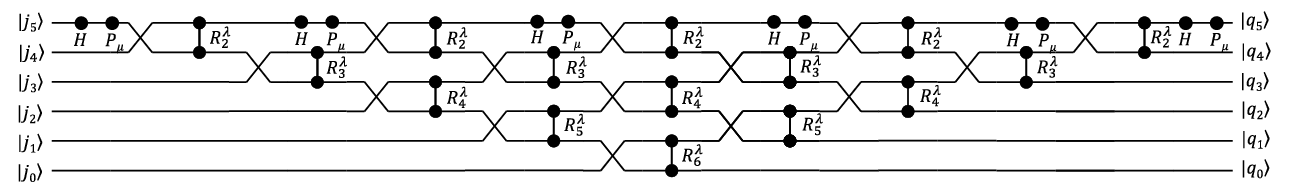, width=14cm}}
	\vspace*{13pt}
	\caption{\label{fig:mQFTQuditsLocal} The modified QFT circuit with local interactions on 6 qudits with parameter $\lambda$. Single qudit Hadamard gates $H^{(p)}$ and permutation $P_{\mu}^{(p)}$ gates are denoted with a bullet, while two-qudit rotation gates $R_{k}^{\lambda(p)}$ are denoted with two bullets connected with a vertical line. Swap gates are denoted with crossing lines. Note that the qudits are rearranged in the correct reverse order at the end of the computation. }
\end{figure}

\begin{figure}[!h]
	\centering{\epsfig{file=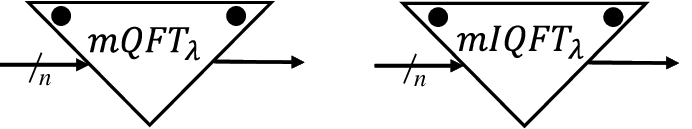, width=8cm}}
	\vspace*{13pt}
	\caption{\label{fig:mQFTlocalSymbol} Symbols for $mQFT_{\lambda}^{(p)}$ and its inverse $mIQFT_{\lambda}^{(p)}$ with local interactions Thick bullets at the input and ouput sides denote the most significant qudit positions. }
\end{figure}

\noindent
The near-identical structure of the $mQFT_{\lambda}^{(p)}$ circuit with the QFT one permits its implementation in 1D-LNN architectures using a similar structure of Fig. \ref{fig:QFTQuditsLocal}. In this case a permutation gate $P_{\mu}^{(p)}=P_{\lambda ^{-1}}^{(p)}$ is added after each Hadamard gate while the rotation gates are replaced with $R_{k}^{(p)\lambda}$ gates as shown in Figure \ref{fig:mQFTQuditsLocal}. Note that in this structure the qudits appear at the end at the correct reverse order, similarly to the normal QFT circuit of Figure \ref{fig:QFTQuditsLocal}. Like the normal QFT case, the depth  is $2n-1$ steps (added permutation gates are executed concurrently with the SWAP gates) and the quantum cost is again $n(n+1)/2$ in terms of these gates. Consequently, the full constant multiplier has twice the depth and the cost of each one of its consituent parts, that is depth of $4n-1$ steps and gate cost $n(n+1)$. Figure \ref{fig:mQFTlocalSymbol} depicts the symbols corresponding to the local interactions circuits of  $mQFT_{\lambda}^{(p)}$ and its inverse.

The are three approaches to design the inverse of a constant multiplier. The first two are derived from the relation
\begin{equation}\label{eq:INVMODMULCQudits1}
MODMULC_{\lambda}^{(p)-1}= [ QFT^{(p)-1}\cdot mQFT_{\lambda}^{(p)}]^{-1}=  mQFT_{\lambda}^{(p)-1}\cdot QFT^{(p)}    
\end{equation}

\noindent
The topology of inverse modified QFT in the above relation, is just a horizontally flipped (left-right) circuit $mQFT_{\lambda}^{(p)}$, whose controlled rotation gates have the opposite angles and the permutation gates $P_{\lambda ^{-1}}^{(p)}$ are replaced by their inverses $P_{\lambda}^{(p)}$ (first choice). Another construction (second choice) for the $mQFT_{\lambda}^{(p)-1}$ comes from the observation  

\begin{equation}\label{eq:mQFTinv}
mQFT_{\lambda}^{(p)-1}=mQFT_{-\lambda}^{(p)}
\end{equation} 

\noindent 
which is derived from  Eq. \eqref{eq:mQFTqudits}.

The third choice to construct an inverse  modified QFT emerges by observing that $MODMULC_{\lambda ^{-1}}^{(p)} \cdot MODMULC_{\lambda}^{(p)} \ket{x} \ket{\lambda^{-1}\lambda x \bmod p^{n} }=  \ket{x},\, \forall x$. Thus, the inverse multiplier of parameter $\lambda$ is a direct multiplier with parameter $\lambda^{-1}$ (third choice); 

\begin{equation}\label{eq:INVMODMULCQudits2}
MODMULC_{\lambda}^{(p)-1}=MODMULC_{\lambda ^{-1}}^{(p)}    
\end{equation}
Finally, it is useful to find which is the element $A\in SL_{2}(\mathbb{Z}_N)$ giving as unitary representation the $U(A)_{k,l}=mQFT_{\lambda}^{p}$. Using \eqref{eq:IQFTmQFTopQudits} we derive $mQFT_{\lambda}^{(p)} = QFT^{(p)} \cdot MODMULC_{\lambda}^{p} $ and from the represenations \eqref{Uepsilon} and \eqref{Umul} we get

\begin{equation}
 mQFT_{\lambda}^{(p)} =U \left( \begin{array}{cc} 0 & -\lambda^{-1} \\ \lambda & 0 \end{array} \right) 
\label{UmQFT}
\end{equation}

\section{Diagonal operators on multilevel qudits}\label{sec:DiagQudits}

\subsection{Quadratic diagonal operator}\label{subsec:SqDiagQudits}

\noindent
Representaion of Eq. \eqref{Udiag} leads to the requirement of a diagonal operator with quadratic phases multiplied by a contant integer. We begin with the following operator definition on $n$ qudits of $p$ dimensions with 

\begin{equation}\label{eq:SqDiagDefQudits}
\Delta _{sq}^{(p)} \doteq \sum_{x=0}^{p^{n}-1} e^{i\frac{2\pi}{p^{n}}x^{2}} \op{x}{x} = \sum_{x=0}^{p^{n}-1} e^{i\frac{2\pi}{p^{n}}(x^{2} \bmod p^{n})} \op{x}{x}
\end{equation}

\noindent
In the $p$-ary representation $x=(x_{n-1}\ldots x_{1}x_{0})$ we can express  $x^{2} \bmod p^{n}$ as

\begin{equation}\label{eq:xSquareD}
x^{2}\bmod p^{n}=\sum_{j=0}^{n-1}\sum_{l=j}^{n-1} x_{j}x_{l-j}p^{l} \bmod p^{n} = \sum_{l=0}^{n-1}\sum_{j=0}^{l} x_{j}x_{l-j}p^{l} \bmod p^{n}
\end{equation}

\noindent
Introducing the second equality of Eq. \eqref{eq:xSquareD} into Eq.  \eqref{eq:SqDiagDefQudits} and decomposing $\ket{x}=\ket{x_{n-1}}\cdots \ket{x_{1}}\ket{x_{0}}$ we derive

\begin{equation}\label{eq:SqDiagDerivationQudits}
\begin{split}
\Delta _{sq}^{(p)} = & \sum_{x_{n-1}=0}^{p-1} \cdots \sum_{x_{1}=0}^{p-1} \sum_{x_{0}=0}^{p-1}  e^{i\frac{2\pi}{p^{n}}\sum_{l=0}^{n-1}\sum_{j=0}^{l} x_{j}x_{l-j}p^{l}} \op{x_{n-1}}{x_{n-1}}\cdots \op{x_{1}}{x_{1}} \cdot \op{x_{0}}{x_{0}} = \\
& \sum_{x_{n-1}=0}^{p-1} \cdots \sum_{x_{1}=0}^{p-1} \sum_{x_{0}=0}^{p-1}  \prod_{l=0}^{n-1}\prod_{j=0}^{l} e^{i\frac{2\pi}{p^{n-l}} x_{j}x_{l-j}} \bigotimes _{m=n-1}^{0} \op{x_{m}}{x_{m}}
\end{split}
\end{equation}

\noindent
Because $\op{x}{x},\, \forall x \in{0\ldots p^{n}-1}$ are orthogonal projectors, it holds that 

\begin{equation}\label{eq:projectorPropertyD}
\sum_{x=0}^{p^{n}-1}\prod_{k=1}^{c} f_{k}(x) \op{x}{x} = \prod_{k=1}^{c} \sum_{x=0}^{p^{n}-1} f_{k}(x) \op{x}{x} 
\end{equation}

\noindent
Thus, Eq. \eqref{eq:SqDiagDerivationQudits} becomes

\begin{equation}\label{eq:SqDiagDerivationFinalQudits}
\begin{split}
\Delta _{sq} = & \prod_{l=0}^{n-1}\prod_{j=0}^{l} \left( \sum_{x_{n-1}=0}^{p-1} \cdots \sum_{x_{1}=0}^{p-1} \sum_{x_{0}=0}^{p-1}   e^{i\frac{2\pi}{p^{n-l}} x_{j}x_{l-j}} \bigotimes _{m=n-1}^{0} \op{x_{m}}{x_{m}} \right) = \\
& \prod_{l=0}^{n-1}\prod_{j=0}^{l} R_{n-l}^{(p)(j,l-j)}
\end{split}
\end{equation}
The factors inside the parenthesis, denoted as $R_{n-l}^{(p)(j,l-j)}$, are exactly the  rotation gates of Eqns. \eqref{eq:Rkfull} and \eqref{eq:singleRkfull}. In the case where $j\neq l-j$ the gates $R_{n-l}^{(p)(j,l-j)}$ correspond to the two qudit rotation gates of Eq.\eqref{eq:Rkfull} controlled by the $j$ qudit and targeting the $(l-j)$ qudit. In the case where $j=l-j$, the gates  $R_{n-l}^{(p)(j,l-j)}$ are the single qudit gates of Eq. \eqref{eq:singleRkfull} applied at qudit $j=l/2$ .  Thus,  Eq. \eqref{eq:SqDiagDerivationFinalQudits}  directly describes the circuit topology of the diagonal operator $\Delta _{sq}^{(p)}$, which is given in Figure \ref{fig:sqDiagQudits} for the case of $n=4$ qudits. Each partial dit product $x_{j}x_{l-j}$ corresponds to a rotation gate connected between qudits $j$ and $l-j$ (when $j=l-j$ then a single qudit rotation gate is applied on qudit $j=l/2$). The angle parameter of each gate is $2\pi/p^{n-l}$.

\renewcommand{\rgate}[1]{*=<2.3em>[Fo]{#1} \qw} 
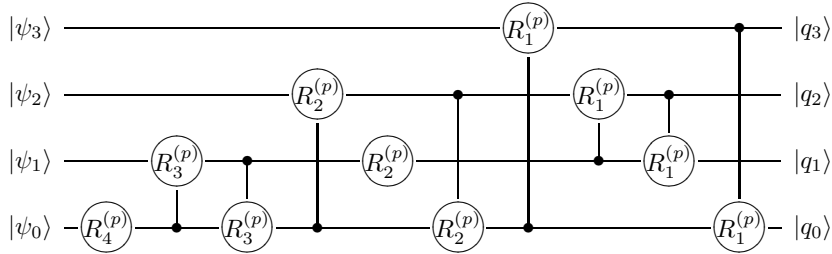
\begin{figure}[!h]
	\scalebox{.95}{\centerline{
			\Qcircuit @C=0.5em @R=0.35em {
				\lstick{\ket{\psi _{3}}}   & \qw			& \qw		    & \qw			& \qw			& \qw				& \qw		   & \rgate{R_{1}^{(p)}}	& \qw			& \qw			& \ctrl{3}	& \rstick{\ket{q_{3}}} \qw \\
				\lstick{\ket{\psi _{2}}}   & \qw			& \qw			& \qw			& \rgate{R_{2}^{(p)}}	& \qw				& \ctrl{2}	   & \qw		 & \rgate{R_{1}^{(p)}} & \ctrl{1}		    & \qw			& \rstick{\ket{q_{2}}} \qw \\
				\lstick{\ket{\psi _{1}}}   & \qw			& \rgate{R_{3}^{(p)}}	& \ctrl{1}		& \qw			 & \rgate{R_{2}^{(p)}}	& \qw			& \qw 	& \ctrl{-1}		& \rgate{R_{1}^{(p)}} 		& \qw 			& \rstick{\ket{q_{1}}} \qw \\
				\lstick{\ket{\psi _{0}}}   & \rgate{R_{4}^{(p)}} 	& \ctrl{-1}		& \rgate{R_{3}^{(p)}}	& \ctrl{-2}		 &  \qw				& \rgate{R_{2}^{(p)}}& \ctrl{-3}			& \qw		& \qw	    	& \rgate{R_{1}^{(p)}}		& \rstick{\ket{q_{0}}} \qw \\
			}
		}
	}
	\vspace*{13pt}
	\caption{\label{fig:sqDiagQudits} Quadratic diagonal circuit on $4$ qudits as derived by Eq. \eqref{eq:SqDiagDerivationFinalQudits}.}
\end{figure} 

\noindent
The diagonal circuit of Figure \ref{fig:sqDiagQudits} adopts several simplifications which can lead to lower quantum cost, improved depth and local interactions capability. First, we can use the fact that all the involved gates are diagonal, so they mutually commute. Second, we observe that the control and target qudits of each gate can be interchanged due to the identity $R_{n-l}^{(p)(j,l-j)}=R_{n-l}^{(p)(l-j,j)}$. 
It is evident that adjacent gates applied on the same qudits can be merged together defining the new gate $S_{k}^{(p)} \doteq R_{k}^{(p)}\cdot R_{k}^{(p)}=R_{k}^{(p)2}$, The symbol $S_{k}^{(p)}$ stands for "squared" $R_{k}^{(p)2}$ gate and it is also a diagonal gate with two-fold angles with respect to $R_{k}^{(p)}$ . Using this notation, we have a simplified circuit like that of Figure \ref{fig:sqDiagQudits3}. 

\renewcommand{\rgate}[1]{*=<2.3em>[Fo]{#1} \qw} 
\begin{figure}[!h]
	\scalebox{.95}{\centerline{
			\Qcircuit @!C=0.5em @!R=0.35em {
				\lstick{\ket{\psi _{3}}}   & \qw			& \qw		    & \qw			& \qw				& \qw	& \rgate{S_{1}^{(p)}} 	 & \rstick{\ket{q_{3}}} \qw \\
				\lstick{\ket{\psi _{2}}}   & \qw			& \qw			& \qw			& \rgate{S_{2}^{(p)}} 	& \rgate{S_{1}^{(p)}}  & \qw &\rstick{\ket{q_{2}}} \qw \\
				\lstick{\ket{\psi _{1}}}   & \qw			& \rgate{S_{3}^{(p)}}	& \rgate{R_{2}^{(p)}} & \qw 			& \ctrl{-1} 	& \qw & \rstick{\ket{q_{1}}} \qw \\
				\lstick{\ket{\psi _{0}}}   & \rgate{R_{4}^{(p)}} 	& \ctrl{-1}		& \qw			& \ctrl{-2} 	&	\qw 	& \ctrl{-3} & \rstick{\ket{q_{0}}} \qw \\
			}
		}
	}
	\vspace*{13pt}
	\caption{\label{fig:sqDiagQudits3} Simplified quadratic diagonal circuit on $4$ qudits.}
\end{figure}
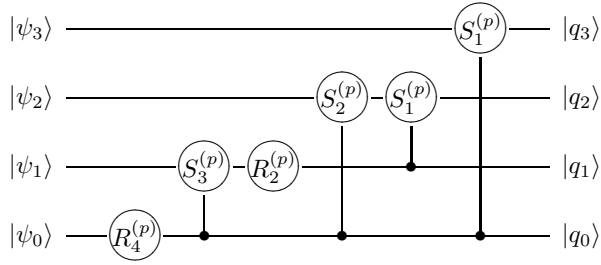 
\noindent
The above described simplifications can be expressed rigorously by suitably modifying  Eq. \eqref{eq:SqDiagDerivationFinalQudits} as follows: 

\begin{equation}\label{eq:SqDiagDerivationSimpler1Qudits}
\Delta _{sq}^{(p)} = \prod_{l=0}^{n-1} \left( \prod_{j=0}^{\floor{\frac{l+1}{2}}-1} R_{n-l}^{(p)(j,l-j)} \right) \left( \prod_{j=\floor{\frac{l+1}{2}}}^{\ceil{\frac{l+1}{2}}-1} R_{n-l}^{(p)(j,l-j)} \right) \left( \prod_{j=\ceil{\frac{l+1}{2}}}^{l} R_{n-l}^{(p)(j,l-j)} \right)
\end{equation}

\noindent
The middle factor of the inner product (the one with limits $j=\floor{\frac{l+1}{2}}\ldots \ceil{\frac{l+1}{2}}-1 $ is identity for odd $l$, otherwise, for even $l$, it equals to $R_{n-l}^{(p)(l/2,l/2)}$ and corresponds to the single qudit rotation gates of Figures \ref{fig:sqDiagQudits}  and \ref{fig:sqDiagQudits3}.

\noindent
The last factor of inner loop in Eq. \eqref{eq:SqDiagDerivationSimpler1Qudits}  can be re-expressed as 
\begin{equation}\label{eq:SqDiagDerivationSimpler2Qudits}
\left( \prod_{j=\ceil{\frac{l+1}{2}}}^{l} R_{n-l}^{(p)(j,l-j)} \right) = \left( \prod_{j'=0}^{l-\ceil{\frac{l+1}{2}}} R_{n-l}^{(p)(l-j',j')} \right) = \left( \prod_{j'=0}^{\floor{\frac{l+1}{2}}-1} R_{n-l}^{(p)(j',l-j')} \right)
\end{equation}
\noindent 
where the index changing $j'=l-j$ has been applied. Then Eq. \eqref{eq:SqDiagDerivationSimpler1Qudits} is equivalent to

\begin{equation}\label{eq:SqDiagDerivationSimpler3Qudits}
\begin{split}
\Delta _{sq}^{(p)} = & \prod_{l=0}^{n-1} \left( \prod_{j=0}^{\floor{\frac{l+1}{2}}-1} R_{n-l}^{(p)(j,l-j)} \right)^{2} \left( \prod_{j=\floor{\frac{l+1}{2}}}^{\ceil{\frac{l+1}{2}}-1} R_{n-l}^{(p)(j,l-j)} \right) = \\
& \prod_{l=0}^{n-1} \left( \prod_{j=0}^{\floor{\frac{l+1}{2}}-1} S_{n-l}^{(p)(j,l-j)} \right) \left( \prod_{j=\floor{\frac{l+1}{2}}}^{\ceil{\frac{l+1}{2}}-1} R_{n-l}^{(p)(j,l-j)} \right)
\end{split}
\end{equation}
The form of Eq. \eqref{eq:SqDiagDerivationSimpler3Qudits} shows that for each $l$, all the gates involved in the inner loop can be executed concurrently, if the architecture permits such an operation. Concurrent operation of the gates  is feasible because $j=0\ldots \floor{\frac{l+1}{2}}-1 \implies j\neq l-j, \forall l$. Thus, all the gates of the left parenthesis in Eq. \eqref{eq:SqDiagDerivationSimpler3Qudits} are applied on different control $j$ and target qudits $l-j$. Moreover, right parenthesis of  Eq. \eqref{eq:SqDiagDerivationSimpler3Qudits} corresponds to $R_{n-l}^{(p)(l/2,l/2)}$ gate for even $l$, and thus it can be executed concurrently, too. Algorithm \ref{table:sqDiagSynthesisD} is the reformulation of Eq. \eqref{eq:SqDiagDerivationSimpler3Qudits} explicitly showing the synthesis of the quadratic diagonal operator in a parallel execution mode. Outer loop index $l$ runs up to $n-1$ and all the gates inside this loop are executed in parallel, thus the  depth of the derived circuit is $n$ while the number of $R_{(p)}$ and $S_{(p)}$ gates involved is about $n^2/2$.

\begin{algorithm}[!h]
	\SetAlgoNoLine
	\For{$l=0\ldots n-1$}{
		\bfseries{in parallel do} \\ 
		\For{$j=0\ldots \floor{\frac{l+1}{2}}-1$ \bf{in parallel}} {
			$S_{n-l}^{(p)(j,l-j)}$ 
		}
		\If{$l \bmod 2==0$}{
			$R_{n-l}^{(p)(l/2,l/2)}$
		}
	}
	\caption{Synthesis algorithm of the quadratic diagonal operator $\Delta _{sq}^{(p)}$ with parallel execution of gates.}
	\label{table:sqDiagSynthesisD}		
\end{algorithm}

\begin{figure}[!h]
	\centering{\epsfig{file=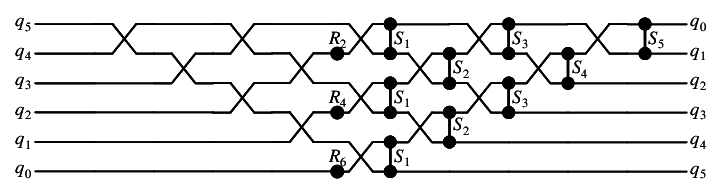, width=8cm}}
	\vspace*{13pt}
	\caption{\label{fig:sqDiagLocalD} Squaring diagonal circuit with local interactions on 6 qudits. Single qudit rotation gates $R^{(d)}$ are denoted with a bullet, while two qudit rotation gates $S^{(d)}$ are denoted with two bullets connected with a vertical line. Swap gates are denoted with crossing lines.  Note that the qudits order is reversed in the end likewise the QFT and modified QFT case, but this order is not correct.}
\end{figure}

\begin{figure}[!h]
	\centering{\epsfig{file=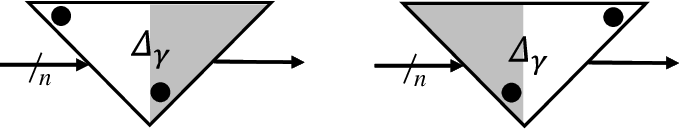, width=6cm}}
	\vspace*{13pt}
	\caption{\label{fig:sqDiagLocaSymbol} Symbols of the quadratic diagonal circuit with local interactions. The triangle symbol denotes the space-time distribution of SWAP and rotation gates  in Figure \ref{fig:sqDiagLocalD}. Left symbol (a) corresponds to the configuration of Figure \ref{fig:sqDiagLocalD} while thr right one (b) corresponds to the flipped left-right configuration. Bullets mark the position of the most significant qudit of the input-output states. Grayed area shows the placement of the rotation gates.}
\end{figure}

Generalization of the simplified topology in Figure \ref{fig:sqDiagQudits3} to larger values of $n$ is not visually straightforward. Figure \ref{fig:sqDiagQudits4} depicts  the case for $n=6$ directly derived by  algorithm \ref{table:sqDiagSynthesisD}. Gates are grouped in steps $l=0\ldots 5$ showing their parallel execution. This topology closely resembles that of a QFT, with the following differences: Single rotation gates of the half lower qudits in Figure \ref{fig:sqDiagQudits4} correspond to the Hadamard gates of the QFT while different angles are used in all the two-qudit rotation gates. Also, in this circuit configuration the order of the qudits is correct at the end, unlike the QFT case. It is easy to map such a topology to a local interactions architecture without significant overhead. Figure \ref{fig:sqDiagLocalD} shows such a mapping for the case $n=6$. Thus, the circuit of the quadratic diagonal operator can be executed on a 1D-LNN machine in linear depth and quadratic quantum gates cost. Namely, the circuit depth of Figure \ref{fig:sqDiagLocalD} is $2n-3$ counted in  two qudit local interactions, while its quantum cost counted in two qudit local interactions is $n^{2}/2$. Observe that in the local interactions configuration the qudits order is reversed at the end, but unlike the QFT case, this order is not the correct one. That is, the most significant qudit of the input state is the top line in Figure  \ref{fig:sqDiagLocalD}, while the most significant qudit of the transformed state is the bottom line. If this circuit is flipped left-right then, because it consists exclusively of rotation gates which mutually commute, its operation is exactly the same with the difference that the most significant input qudit sits at the bottom line, while he most significant output qudit sits at the top line. This fact will be exploited in the construction of the QAFrFT to reduce the overall depth. The symbols for these two configurations are shown as in Figure  \ref{fig:sqDiagLocaSymbol}.

\renewcommand{\rgate}[1]{*=<2.3em>[Fo]{#1} \qw} 
\begin{figure}[!h]
	\scalebox{.95}{\centerline{
			\Qcircuit @!C=0.5em @!R=0.35em {
				\lstick{\ket{\psi _{5}}}   & \qw			&  \qw		    		& \qw			&   \qw					& \qw				& \qw		& \qw &\qw &\qw			& \rgate{S_{1}^{(p)}}				& \qw				& \qw				&  \rstick{\ket{q_{5}}} \qw \\
				\lstick{\ket{\psi _{4}}}   & \qw			&  \qw		    			& \qw			&   \qw					& \qw				& \qw						& \rgate{S_{2}^{(p)}}		& \qw				& \qw		& \qw &\rgate{S_{1}^{(d)}} & \qw		&  \rstick{\ket{q_{4}}} \qw \\
				\lstick{\ket{\psi _{3}}}   & \qw			&  \qw		    			& \qw			&    \qw					& \rgate{S_{3}^{(p)}}			& \qw			& \qw				& \rgate{S_{2}^{(p)}}		& \qw		& \qw &\qw & \rgate{S_{1}^{(p)}}		&\rstick{\ket{q_{3}}} \qw \\
				\lstick{\ket{\psi _{2}}}   & \qw			&  \qw					& \rgate{S_{4}^{(p)}}	&  \qw			 		&    \qw		& \rgate{S_{3}^{(p)}}					& \qw				& \qw				& \rgate{R_{2}^{(p)}}	& \qw &\qw & \ctrl{-1}	&\rstick{\ket{q_{2}}} \qw \\
				\lstick{\ket{\psi _{1}}}   & \qw			&  \rgate{S_{5}^{(p)}}				& \qw			&    \rgate{R_{4}^{(p)}} 	&  \qw		& \ctrl{-1}								&  \qw		& \ctrl{-2}			& \qw		& \qw &\ctrl{-3} & \qw		&\rstick{\ket{q_{1}}} \qw \\
				\lstick{\ket{\psi _{0}}}   & \rgate{R_{6}^{(p)}} 	&  \ctrl{-1}					& \ctrl{-2}			&  \qw		& \ctrl{-3}				& \qw					& \ctrl{-4}			& \qw				& \qw		& \ctrl{-5} &\qw & \qw		&\rstick{\ket{q_{0}}} \qw \\
				\lstick{\textrm{Step}}  & l=0 	  	& l=1			& 	l=2	& 		& 	l=3			& 				& l=4		& 			& 		&l=5 & &		& \\
			}
		}
	}
	\vspace*{13pt}
	\caption{\label{fig:sqDiagQudits4} Quadratic diagonal circuit on $6$ qudits with parallel execution of gates in $6$ steps.}
\end{figure}
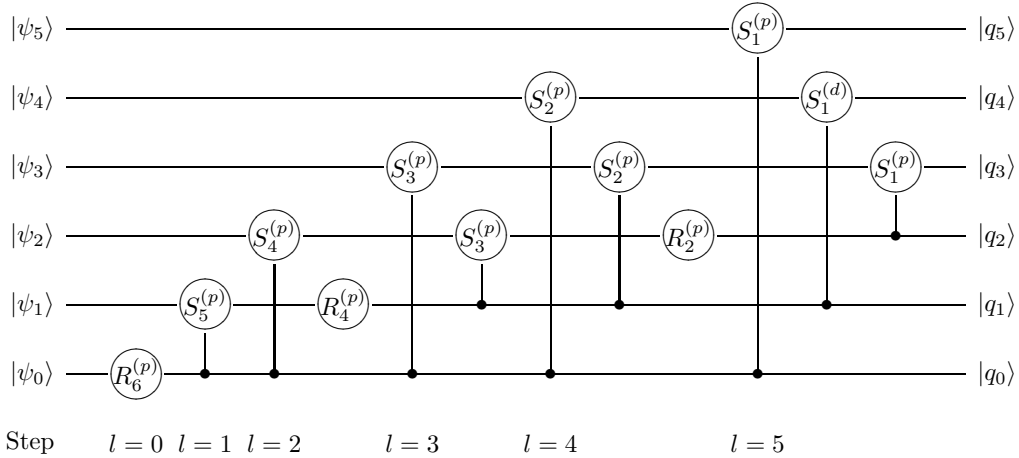

\subsection{Constant factor quadratic diagonal operator}\label{subsec:cSqDiagQudits}

The generalization of the  quadratic diagonal operator required by the AFrFT application, is to permit a constant integer factor $\gamma$ applied in each of the phase angles. This is the constant factor quadratic diagonal operator on $n$ qudits defined by

\begin{equation}\label{eq:cSqDiagDefQudits}
\Delta _{\gamma , sq}^{(p)} \doteq \sum_{x=0}^{p^{n}-1} e^{i\frac{2\pi}{p^{n}}\gamma x^{2}} \op{x}{x}
\end{equation}

\noindent
This does not differentiate essentially the previous analysis and leads to 

\begin{equation}\label{eq:cSqDiagDerivationFinalQudits}
\Delta _{\gamma , sq}^{(p)} = 
\prod_{l=0}^{n-1}\prod_{j=0}^{l} \left( R_{n-l}^{(p)(j,l-j)} \right)^{\gamma}
\end{equation}

\noindent
which is similar to Eq. \eqref{eq:SqDiagDerivationFinalQudits}, the only difference is that the applied gates $\left( R_{n-l}^{(p)(j,l-j)} \right)^{\gamma}$  controlled by the $j$ qudit and targeting the $(l-j)$ qudit have different angle parameters compared to the ones of Eq. \eqref{eq:SqDiagDerivationFinalQudits}. These are modified rotation gates like the ones used in the modified QFT of Section \ref{sec:MultQudits}. Except of this difference, the simplification, topologies, depths and quantum costs of the constant factor quadratic diagonal operator are exactly the same with the ones of the simple quadratic diagonal operator discussed in the previous subsection \ref{subsec:SqDiagQudits}.

\section{Quantum Arithmetic Fractional Fourier Transform circuit}\label{sec:QFrFTcircuit}

The construction of the QFT based multiplier of Section  \ref{sec:MultQudits} and the constant factor quadratic diagonal operator of subsection \ref{subsec:cSqDiagQudits} together with the normal QFT permits the implementation of the Quantum Arithmetic Fractional Fourier Transform as defined in Section \ref{sec:FQMandFrFT}. The QAFrFT on $p^{n}$ dimensions with parameters $a$ and $b$ as defined in Eq. \eqref{AFrFT} can be expressed using Eq. \eqref{FrFTgroupDecomposition} decomposition and the representations of Eqns. \eqref{Umul}, \eqref{Udiag}, \eqref{Uepsilon} and \eqref{UmQFT} as

\begin{equation}\label{eq:QFrFTdecomposition}
\begin{split}
QAFrFT= & \Delta_{-a/(2b) ,sq}^{(p)}\cdot MODMULC_{b}^{(p)} \cdot QFT^{(p)} \cdot \Delta_{-a/(2b) ,sq}^{(p)} =\\
& \Delta_{-a/(2b) ,sq}^{(p)}  \cdot mQFT_{b^{-1}}^{(p)}  \cdot \Delta_{-a/(2b) ,sq}^{(p)}
\end{split}
\end{equation} 

\noindent
The operator $\Delta_{-a/(2b) ,sq}^{(p)}$ is the operator of Eq. \eqref{eq:cSqDiagDefQudits} on $n$ qudits with $\gamma=-a/(2b) \,(\bmod \, \, p^{n})$, the QFT operators are defined on $n$ qudits and the modified QFT operator $mQFT_{b^{-1}}^{(p)}$ is that of  Eq. \eqref{eq:mQFTqudits} with $\lambda=b^{-1}$. The last part of Eq. \eqref{eq:QFrFTdecomposition} is derived by observing that $MODMULC_{b}^{(p)} \cdot QFT^{(p)}$ is the representation of

\begin{equation}\label{eq:MULxQFT}
\left( \begin{array}{cc}
b & 0 \\
0 & b^{-1}
\end{array} \right) \cdot
\left( \begin{array}{cc}
0 & -1 \\
1 & 0
\end{array} \right) =
\left( \begin{array}{cc}
0 & -b \\
b^{-1} & 0
\end{array} \right)
\end{equation}

\noindent
and taking into account the representation of Eq. \eqref{UmQFT}.

\begin{figure}[!h]
	\centering{\epsfig{file=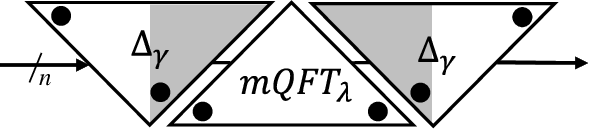, width=6cm}}
	\vspace*{13pt}
	\caption{\label{fig:QFrFTBlock} Block diagram of Quantum Arithmetic Fractional Fourier Transform Circuit on qudits of dimension $p$. The blocks used are  two  constant quadratic diagonal operators $\Delta_{\gamma ,sq}^{(d)}$ with parameter $\gamma=-a/(2b) \pmod{p^{n}} $ and a modified QFT circuit with parameter $\lambda=b^{-1}$. The right diagonal operator is the left one but flipped left-right. Bullets denote the most significant qudit evolution in space.}
\end{figure} 
This sequence of operators is depicted  in the block diagram of Figure \ref{fig:QFrFTBlock}. The triangular shape of the blocks reflects the space-time distribution of the gates as shown in  Figures \ref{fig:mQFTQuditsLocal} and \ref{fig:sqDiagLocalD}, of the modified QFT and the constant factor quadratic operator, respectively.  Because the first diagonal block reverses the qudits order  at the end of the computation, the middle mQFT block is flipped upside-down. The last diagonal operator is flipped left-right so as to accept the most significan qudit at its botom line, as explained in subsection \ref{subsec:SqDiagQudits}. This configuration implies that while computations gradually seaze at the qudits of one block they can gradually initiated in the next block. The depth of each triangular block is about $2n$ (counted in two qudits interactions of adjacent swap and rotation gates), thus the overall depth of the circuit in Figure \ref{fig:QFrFTBlock} is about $4n$ (The mQFT block is essentially applied when the qudits would remain idle in between the two diagonal blocks). Given that the quantum cost of each trangular block is about $n^{2}/2$ in single and two qudit gates (where adjacent SWAP and rotation gates are counted as merged), the total quantum cost of the quantum AFrFT is about $1.5n^{2}$.

In Appendix A we give an example of the AFrFT matrix which corresponds to a generator element of $SO_{2}(\mathbb{Z}_{11})$ along with its decomposition. 

\section{Variations of the proposed quantum circuits}\label{sec:Variations}
\begin{figure}[!h]
	\centering{\epsfig{file=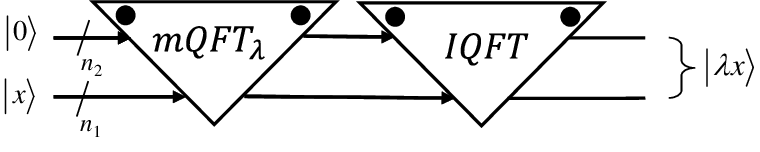, width=8cm}}
	\vspace*{13pt}
	\caption{\label{fig:MultOddBlockQudits} The constant multiplier $MULC_{\lambda}^{(p)}$ for $\gcd(\lambda,p^{n})=1$ consisting of a modified QFT and an inverse QFT on $n=n_{1}+n_{2}$ qudits, in succession. The initial state of the $n_{2}=\ceil{\log_{p}\lambda}$ most significant qubits is zero, while the input argument $\ket{x}$ is fed on the $n_{1}$ least significant qudits.}
\end{figure}

In this Section some variations of the two proposed quantum circuits (multiplier and diagonal operator) are given, which they have a broader application. Also, it is shown how they can be adapted in two-level qubits architecture, althought the QAFrFT itself can not be adapted easily in this context.
In the case one needs a normal multiplier for constants co-prime with $p^{n}$ on qudits, it is straightforward to use the  constant modulo $p^{n}$ multiplier $MODMULC_{\lambda}^{(p)}$ as a normal one, retaining the restriction $\gcd(\lambda ,p^{n})=1$. Suppose our requirement is the multiplication of integer states $\ket{x}$ of $n_{1}$ qudits in the range $\ket{0}\ldots \ket{p^{n_{1}}-1}$ by a constant $\lambda$  which fits into $n_{2}$ dits ($p$-ary digits), that is $n_{2}=\ceil{\log _{p}\lambda}$. In this case the maximum possible product  $\ket{\lambda x}$ fits in $n=n_{1}+n_{2}$ qudits and it holds that $\ket{\lambda x} = \ket{\lambda x \bmod p^{n}}$. Consequently this multiplier can be designed as a  modulo $p^{n}$ multiplier of $n=n_{1}+n_{2}$ qudits whose lower $n_{1}$ qudits are fed with the argument $\ket{x}$ while its upper $n_{2}$ qudits are initially in the zero state $\ket{0}$. This construction is shown in Figure \ref{fig:MultOddBlockQudits}.

\begin{figure}[!h]
	\centering{\epsfig{file=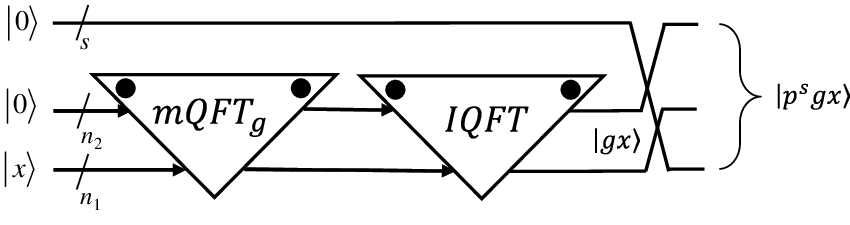, width=8cm}}
	\vspace*{13pt}
	\caption{\label{fig:MultBlockQudits} Constant multiplier $MULC_{\lambda}$ for any constant $\lambda=g\cdot p^{s}$ with $\gcd(g,p^{n})=1$. The initial state of the $n_{2}$ most significant qudits is zero, while the input argument $\ket{x}$ is fed in its $n_{1}$ least significant qudits. After the application of the $mQFT_{g}^{(p)}$ and $QFT^{(p)-1}$, all the qudits are "left" rotated (up in this figure) by $s$ positions.}
\end{figure}

Next we further extend the previous normal multiplier waiving the requirement that constant $\lambda$ must be co-prime with $p^{n}$. That is, $\lambda$ can be any integer. In this case, the fundamental theorem of the arithmetic states that $\lambda$ can be factorized as  $\lambda=g\cdot  p^{s}$ where $\gcd(g,p^n)=1$ and $0<s<n=n_{1}+n_{2}$ (Again, $n_{1}$ are the number of qudits holding the mutltiplcand and $n_{2}=\ceil{\log_{p}\lambda}$).  Then, the classical multiplication by $\lambda$ is reduced to multiplication by the constant $g$ and a left shift by $s$ dits. The respective quantum circuit has to multiply $g$, which is of $n_{2}$ dits, with the input argument of $n_{1}$ qudits. As the result fits in $n_{1}+n_{2}$ qudits, we apply the previous multiplier of Figure \ref{fig:MultOddBlockQudits} of width   $n_{1}+n_{2}$ qudits, feed its upper  $n_{2}$ qudits with the zero state and its lower $n_{1}$ qudits with the input state $\ket{x}$. Another register of $s$ qudits initially in the zero state is also used to perform a "left" rotation of the multiplier product together with these $s$ qudits as shown in Figure \ref{fig:MultBlockQudits}.

The multiplier circuits presented herein can be adapted to operate on two dimensional qubits. 
For $p=2$ and odd multiplication constant $\lambda$ the condition $\gcd(\lambda,2^{n})=1$ holds, thus the topology of the circuit in Figure \ref{fig:MultOddBlockQudits} can be used. In this case we use the usual Hadamard gates $H$, the controlled $z$-axis rotation gates  $R_{k}=\diag (1,1,1,e^{i2\pi/2^{k}})$ and their modifications $R_{k}^{\lambda}=\diag (1,1,1,e^{i2\pi \lambda/2^{k}})$ inside the mQFT and QFT blocks. The permutation gate $P_{\mu}^{(2)}$ is not required in the mQFT block because for  odd $\lambda$ it degenerates to the identity gate. If the constant $\lambda$ is even then,  the topology of Figure \ref{fig:MultBlockQudits} can be directly applied in the qubits case as the constant $\lambda = g2^{s}$  with $g$ odd.

Finally, the constant factor quadratic diagonal operator can be easily adapted on two-dimensional qubits as there is no restriction on the value of the factor $\gamma$. For $p=2$, the gate $S_{k} =R_{k}^{2}$ used in the diagonal operator is just the gate $R_{k-1}$. In this case, the gates $S_{1}=R_{0}$ of the last step in Figure \ref{fig:sqDiagQudits4}  are the identity gates, thus the top-most qubit has no applied gates.

\section{Conclusions}\label{sec:Conclusions}
In this work we have presented a new definition of the discrete fractional Fourier tranform operating on the Hilbert space of  $n$ qudits and dimension $p^n$ for $p$ prime. The construction is based on the geometry of the discrete rotation group  $SO_{2}[\mathbb{Z}_p^n]$ acting on the toroidal phase-space $\mathbb{Z}_N \times \mathbb{Z}_N$. Due to the $\bmod p^n$ operation the elements of this group produce random motion in the phase-space. Accordingly the quantization of the generator of this group, which is our definition of the quantum  arithmetic fractional Fourier transform (QAFrFT) following the rules of FQM, produces a chaotic unitary matrix with matrix elements $p^n$ roots of unity.
A Gaussian wave packet in the finite Hilbert space will be randomized after a few operations of the AFrFT. This immediately invites for applications to cryptography methods which exploit chaotic codes \cite{Akhshani:2014}.

Subsequently in this work we decomposed the AFrFT unitary matrix into multiplication quantum circuits, QFT circuits and new quantum circuits for diagonal operators with quadratic phases, called chirps in the signal processing community.
Moreover, the multipliers and  diagonal circuits presented herein, can be adapted to operate on two dimensional qubits. Further work is in progress for the excecution of the QAFrFT on exisiting quantum platforms for qubits.

The whole quantum fractional Fourier transform circuit uses local interactions only; it is given in a form of product of tensor products of single and two qudit gates, so it is suitable for 1D-LNN architecures. 
It is remarkable the fact that the apparent complexity of the QAFrFT as defined in Eq. \eqref{AFrFT} compared to the QFT definition of Eq. \eqref{eq:QFTqudits} leads to only two-fold depth increase in depth ($4n$ for QAFrFT vs. $2n$ for QFT) and three-fold increase in quantum cost ($1.5n^{2}$ for QAFrFT vs. $0.5n^{2}$ for QFT) on an 1D-LNN architecture capable of parallel execution of gates.
This accomplishment is mainly due to the decomposition of Eq. \ref{eq:QFrFTdecomposition} which permitted the contruction of simpler circuits. The introduction of the $mQFT$ operator and the fact that the $mQFT$ and the diagonal operators have a QFT-like structure led to this result. The above complexity measures are referred to the basic gates used $H^{(p)}$,$P_{\mu}^{p}$, and $R_{k}^{p}$ which operate on their whole dimensional space. To estimate these complexities to a more detailed level we have to take into account their decomposition down to more elementary two-level qudits \cite{Di:2013,Pavlidis:2021}. The decomposition of an $R_{k}^{p}$ requires about $4p^{2}$ elementary gates, while $P_{\mu}^{d}$ gate used in the $mQFT$ requires $p$ elementary gates (for $p\neq 2$). With these considerations we have the following Table concerning the QAFrFT  and its constituent circuits.

\begin{table}[!h]
	\caption{\label{table:Complexity} Quantum cost, depth and width of the proposed arithmetic circuits.}
	\centerline{ %\smalllineskip %\footnotesize
			\begin{tabular}{l c c c}
				\hline
				Circuit 	& Cost  			& Depth 		& Width \\\hline
				QAFrFT 		& $6p^{2}n^{2}$		& $16p^{2}n$	& $n$  \\
				mQFT 		& $2p^{2}n^{2}$ 	& $8p^{2}n$ 	& $n$ \\
				MODMULC 	& $4p^{2}n^{2}$ 	& $16p^{2}n$	& $n$ \\
				Diagonal 	& $2p^{2}n^{2}$ 	& $8p^{2}n$ 	& $n$ \\
				\hline
			\end{tabular}
	}
\end{table}
Another consequence of the decomposition given in Eq. \eqref{eq:QFrFTdecomposition} is that the AFrFT accomodates fast classical computation like the Fast Fourier Transform (FFT). The classical AFrFT computation of size $N=p^n$  normally has a complexity which is $O(N^2)$ but there exists a fast algorithm with complexity $O(N\log _{p}N)$. This is justified by the structure of the $mQFT$ block which is identical to the QFT structure, and the QFT structure leads to the FFT algorithm \cite{Nielsen:2011}. In our case it leads to a radix-$p$ decimation in frequency FFT with a complexity  $O(N\log _{p}N)$. On the other side, the diagonal operators in  Eq. \eqref{eq:QFrFTdecomposition} have a classical complexity $O(N)$, and this leads to a total complexity $O(N\log _{p}N)$  for the classical AFrFT.

%\nonumsection{References}
%\noindent
%\bibliographystyle{IEEEtran} 
%\bibliography{qbib}

%\nocite{*}

%\section*{References}
%\noindent
\appendix
\section{Appendix}\label{appendix:A}

We give explicitly an example of the AFrFT corresponding to one generator element of $SO_{2}(\mathbb{Z}_{11})$ as well as the matrices which correspond to the decomposition given in Eq. \eqref{eq:QFrFTdecomposition}. Such an AFrFT is the $1/3$ fraction of the DFT with dimension $11$. There are four generators of $SO_{2}(\mathbb{Z}_{11})$ and we choose the generator
\begin{equation}
g=\left(\begin{array}{cc}3&5\\-5&3\end{array}\right)
\end{equation}
to define the AFrFT, so from Eq. \eqref{AFrFT} we get for the AFrFT with parameters $a=3$ and $b=-5$ the following matrix expressed in powers of $\omega=e^{i\frac{2\pi}{11}}$

\begin{equation}
AFrFT_{(3,-5)}=\left(
\begin{array}{ccccccccccc}
	1&\omega^{8}&\omega^{10}&\omega^{6}&\omega^{7}&\omega^{2}&\omega^{2}&\omega^{7}&\omega^{6}&\omega^{10}&\omega^{8}\\
	\omega^{8}&\omega^{7}&1&\omega^{9}&\omega&\omega^{9}&1&\omega^{7}&\omega^{8}&\omega^{3}&\omega^{3}\\
	\omega^{10}&1&\omega^{6}&\omega^{6}&1&\omega^{10}&\omega^{3}&\omega&\omega^{4}&\omega&\omega^{3}\\
	\omega^{6}&\omega^{9}&\omega^{6}&\omega^{8}&\omega^{4}&\omega^{5}&1&1&\omega^{5}&\omega^{4}&\omega^{8}\\
	\omega^{7}&\omega&1&\omega^{4}&\omega^{2}&\omega^{5}&\omega^{2}&\omega^{4}&1&\omega&\omega^{7}\\
	\omega^{2}&\omega^{9}&\omega^{10}&\omega^{5}&\omega^{5}&\omega^{10}&\omega^{9}&\omega^{2}&1&\omega^{3}&1\\
	\omega^{2}&1&\omega^{3}&1&\omega^{2}&\omega^{9}&\omega^{10}&\omega^{5}&\omega^{5}&\omega^{10}&\omega^{9}\\
	\omega^{7}&\omega^{7}&\omega&1&\omega^{4}&\omega^{2}&\omega^{5}&\omega^{2}&\omega^{4}&1&\omega\\
	\omega^{6}&\omega^{8}&\omega^{4}&\omega^{5}&1&1&\omega^{5}&\omega^{4}&\omega^{8}&\omega^{6}&\omega^{9}\\
	\omega^{10}&\omega^{3}&\omega&\omega^{4}&\omega&\omega^{3}&\omega^{10}&1&\omega^{6}&\omega^{6}&1\\
	\omega^{8}&\omega^{3}&\omega^{3}&\omega^{8}&\omega^{7}&1&\omega^{9}&\omega&\omega^{9}&1&\omega^{7}\\
\end{array}
\right)
\label{FrFTexample1}
\end{equation}
The quadratic diagonal matrix which corresponds to the decomposition of Eq. \eqref{eq:QFrFTdecomposition} has a constant parameter $\gamma=\frac{-3}{2\cdot(-5)}=8 \bmod 11$ and is given by

\begin{equation}
\Delta_{8,sq}=\left(
\begin{array}{ccccccccccc}
	1&0&0&0&0&0&0&0&0&0&0\\
	0&\omega^{8}&0&0&0&0&0&0&0&0&0\\
	0&0&\omega^{10}&0&0&0&0&0&0&0&0\\
	0&0&0&\omega^{6}&0&0&0&0&0&0&0\\
	0&0&0&0&\omega^{7}&0&0&0&0&0&0\\
	0&0&0&0&0&\omega^{2}&0&0&0&0&0\\
	0&0&0&0&0&0&\omega^{2}&0&0&0&0\\
	0&0&0&0&0&0&0&\omega^{7}&0&0&0\\
	0&0&0&0&0&0&0&0&\omega^{6}&0&0\\
	0&0&0&0&0&0&0&0&0&\omega^{10}&0\\
	0&0&0&0&0&0&0&0&0&0&\omega^{8}\\
\end{array}
\right)
\end{equation}
The multiplier by $b=-5=6 \bmod 11$ matrix is

\begin{equation}
	MODMULC_{6}=\left(
	\begin{array}{ccccccccccc}
1&0&0&0&0&0&0&0&0&0&0\\
0&0&1&0&0&0&0&0&0&0&0\\
0&0&0&0&1&0&0&0&0&0&0\\
0&0&0&0&0&0&1&0&0&0&0\\
0&0&0&0&0&0&0&0&1&0&0\\
0&0&0&0&0&0&0&0&0&0&1\\
0&1&0&0&0&0&0&0&0&0&0\\
0&0&0&1&0&0&0&0&0&0&0\\
0&0&0&0&0&1&0&0&0&0&0\\
0&0&0&0&0&0&0&1&0&0&0\\
0&0&0&0&0&0&0&0&0&1&0\\
	\end{array}
	\right)
\end{equation}
The QFT matrix for $p=11$ is

\begin{equation}
QFT=\left(
\begin{array}{ccccccccccc}
1&1&1&1&1&1&1&1&1&1&1\\
1&\omega&\omega^{2}&\omega^{3}&\omega^{4}&\omega^{5}&\omega^{6}&\omega^{7}&\omega^{8}&\omega^{9}&\omega^{10}\\
1&\omega^{2}&\omega^{4}&\omega^{6}&\omega^{8}&\omega^{10}&\omega&\omega^{3}&\omega^{5}&\omega^{7}&\omega^{9}\\
1&\omega^{3}&\omega^{6}&\omega^{9}&\omega&\omega^{4}&\omega^{7}&\omega^{10}&\omega^{2}&\omega^{5}&\omega^{8}\\
1&\omega^{4}&\omega^{8}&\omega&\omega^{5}&\omega^{9}&\omega^{2}&\omega^{6}&\omega^{10}&\omega^{3}&\omega^{7}\\
1&\omega^{5}&\omega^{10}&\omega^{4}&\omega^{9}&\omega^{3}&\omega^{8}&\omega^{2}&\omega^{7}&\omega&\omega^{6}\\
1&\omega^{6}&\omega&\omega^{7}&\omega^{2}&\omega^{8}&\omega^{3}&\omega^{9}&\omega^{4}&\omega^{10}&\omega^{5}\\
1&\omega^{7}&\omega^{3}&\omega^{10}&\omega^{6}&\omega^{2}&\omega^{9}&\omega^{5}&\omega&\omega^{8}&\omega^{4}\\
1&\omega^{8}&\omega^{5}&\omega^{2}&\omega^{10}&\omega^{7}&\omega^{4}&\omega&\omega^{9}&\omega^{6}&\omega^{3}\\
1&\omega^{9}&\omega^{7}&\omega^{5}&\omega^{3}&\omega&\omega^{10}&\omega^{8}&\omega^{6}&\omega^{4}&\omega^{2}\\
1&\omega^{10}&\omega^{9}&\omega^{8}&\omega^{7}&\omega^{6}&\omega^{5}&\omega^{4}&\omega^{3}&\omega^{2}&\omega\\
\end{array}
\right)
\end{equation}
Finally, the matrix of the modified Fourier transform with parameter $\lambda=b^{-1}=(-5)^{-1}=2 \bmod 11$ which is used in the second part of the decomposition \eqref{eq:QFrFTdecomposition} is given by

\begin{equation}
mQFT_{2}=\left(
\begin{array}{ccccccccccc}
1&1&1&1&1&1&1&1&1&1&1\\
1&\omega^{2}&\omega^{4}&\omega^{6}&\omega^{8}&\omega^{10}&\omega&\omega^{3}&\omega^{5}&\omega^{7}&\omega^{9}\\
1&\omega^{4}&\omega^{8}&\omega&\omega^{5}&\omega^{9}&\omega^{2}&\omega^{6}&\omega^{10}&\omega^{3}&\omega^{7}\\
1&\omega^{6}&\omega&\omega^{7}&\omega^{2}&\omega^{8}&\omega^{3}&\omega^{9}&\omega^{4}&\omega^{10}&\omega^{5}\\
1&\omega^{8}&\omega^{5}&\omega^{2}&\omega^{10}&\omega^{7}&\omega^{4}&\omega&\omega^{9}&\omega^{6}&\omega^{3}\\
1&\omega^{10}&\omega^{9}&\omega^{8}&\omega^{7}&\omega^{6}&\omega^{5}&\omega^{4}&\omega^{3}&\omega^{2}&\omega\\
1&\omega&\omega^{2}&\omega^{3}&\omega^{4}&\omega^{5}&\omega^{6}&\omega^{7}&\omega^{8}&\omega^{9}&\omega^{10}\\
1&\omega^{3}&\omega^{6}&\omega^{9}&\omega&\omega^{4}&\omega^{7}&\omega^{10}&\omega^{2}&\omega^{5}&\omega^{8}\\
1&\omega^{5}&\omega^{10}&\omega^{4}&\omega^{9}&\omega^{3}&\omega^{8}&\omega^{2}&\omega^{7}&\omega&\omega^{6}\\
1&\omega^{7}&\omega^{3}&\omega^{10}&\omega^{6}&\omega^{2}&\omega^{9}&\omega^{5}&\omega&\omega^{8}&\omega^{4}\\
1&\omega^{9}&\omega^{7}&\omega^{5}&\omega^{3}&\omega&\omega^{10}&\omega^{8}&\omega^{6}&\omega^{4}&\omega^{2}\\
\end{array}
\right)
\end{equation}
The above AFrFT example given in Eq. \eqref{FrFTexample1} is the representation of a particular generator of $SO_{2}(\mathbb{Z}_{11})$. In this case $p=11$ and is of the form $4k+3$ with $k=2$. The order of $SO_{2}(\mathbb{Z}_{11})$ is $4(k+1)$ where $k+1=3$ ($11=4\cdot 2 +3$, see Eq. \eqref{order4k3}), thus this particular AFrFT is the $1/3$ power of the QFT matrix of size $p=11$, up to a global phase, that is 

\begin{equation}
AFrFT_{(3,-5)}^{3}=QFT
\label{AFrFTto3}
\end{equation}

In this particular case, the global phase happens to be $1$. In general the global phase belongs to the set $\{\pm 1,\pm i\}$. To be concrete we give the complete expression of the representation corresponding to a generator $\left( \begin{array}{cc}a&-b\\b&a\end{array} \right)$ of the rotation subgroup $SO_{2}(\mathbb{Z}_{N})$, $N=p$ \cite{Balian:1986,Athanasiu:1994}:

\begin{equation}
U( \left(\begin{array}{cc}a&-b\\b&a\end{array}\right) )_{k,l} = (-2b|N)i^q\frac{1}{\sqrt{N}} \omega ^{-\frac{a(k^{2}+l^{2})-2kl}{2b}}= (-2b|p)i^q AFrFT_{(3,-5)}
\label{AFrFT}
\end{equation}
In the above equation, $(-2b|N)$ is the Jacobi symbol and equals 1 if $-2b$ is the  square of an integer $\bmod N$, otherwise it equals -1. Also, $q=0$ if $p=4k+1$, and $q=1$ if $p=4k-1$.

The complete expression (including global phase) of the representation corresponding to the element $\epsilon = \left(\begin{array}{cc}0&-1\\1&0\end{array}\right)$ is given by

\begin{equation}
U(\epsilon)_{k,l} = (-1)^{k+1}i^q \frac{1}{\sqrt{N}} \omega ^{kl} = (-1)^{k+1}i^q F_{k,l}
\end{equation}
With the above remarks Eq. \eqref{AFrFTto3} is fully justified.

An integer power of an AFrFT is also another AFrFT. For example, 
\begin{equation}
AFrFT_{(3,-5)}^{2}=\left(
\begin{array}{ccccccccccc}
1&\omega^{8}&\omega^{10}&\omega^{6}&\omega^{7}&\omega^{2}&\omega^{2}&\omega^{7}&\omega^{6}&\omega^{10}&\omega^{8}\\
\omega^{8}&\omega^{7}&1&\omega^{9}&\omega^{1}&\omega^{9}&1&\omega^{7}&\omega^{8}&\omega^{3}&\omega^{3}\\
\omega^{10}&1&\omega^{6}&\omega^{6}&1&\omega^{10}&\omega^{3}&\omega^{1}&\omega^{4}&\omega^{1}&\omega^{3}\\
\omega^{6}&\omega^{9}&\omega^{6}&\omega^{8}&\omega^{4}&\omega^{5}&1&1&\omega^{5}&\omega^{4}&\omega^{8}\\
\omega^{7}&\omega^{1}&1&\omega^{4}&\omega^{2}&\omega^{5}&\omega^{2}&\omega^{4}&1&\omega^{1}&\omega^{7}\\
\omega^{2}&\omega^{9}&\omega^{10}&\omega^{5}&\omega^{5}&\omega^{10}&\omega^{9}&\omega^{2}&1&\omega^{3}&1\\
\omega^{2}&1&\omega^{3}&1&\omega^{2}&\omega^{9}&\omega^{10}&\omega^{5}&\omega^{5}&\omega^{10}&\omega^{9}\\
\omega^{7}&\omega^{7}&\omega^{1}&1&\omega^{4}&\omega^{2}&\omega^{5}&\omega^{2}&\omega^{4}&1&\omega^{1}\\
\omega^{6}&\omega^{8}&\omega^{4}&\omega^{5}&1&1&\omega^{5}&\omega^{4}&\omega^{8}&\omega^{6}&\omega^{9}\\
\omega^{10}&\omega^{3}&\omega^{1}&\omega^{4}&\omega^{1}&\omega^{3}&\omega^{10}&1&\omega^{6}&\omega^{6}&1\\
\omega^{8}&\omega^{3}&\omega^{3}&\omega^{8}&\omega^{7}&1&\omega^{9}&\omega^{1}&\omega^{9}&1&\omega^{7}\\
\end{array}
\right)
\label{FrFTexample2}
\end{equation}
is a fractional power of QFT as $\left(AFrFT_{(3,-5)}^{2}\right)^{\frac{3}{2}}=\left(\sqrt{AFrFT_{(3,-5)}^{2}}\right)^{3}=QFT$ up to a global phase. Note that due to homeomorphism of the representation, $AFrFT_{(3,-5)}^{2}$ can be directly computed as the representation of $g^2=\left(\begin{array}{cc}6&-8\\8&6\end{array}\right)$ (up to a global phase $-i$), namely

\begin{equation}
AFrFT_{(3,-5)}^{2} = -i \cdot AFrFT_{(6,8)}
\end{equation}
\bibliographystyle{ieeetr}

\bibliography{qbib202408}

\end{document}